\documentclass[screen, acmsmall]{acmart}
\usepackage{syntax}
\usepackage[normalem]{ulem}
\usepackage{framed}
\usepackage{caption}
\usepackage{multicol}
\usepackage{csvsimple}
\usepackage{svg}
\DeclareMathSymbol{\shortminus}{\mathbin}{AMSa}{"39}

\usepackage{array}
\usepackage{datatool}
\usepackage{hyperref}
\usepackage{tikz}
\usepackage{pgfplots}
\usepackage{booktabs}
\usepackage{subcaption}
\usepackage{amsmath}
\usepackage{etoolbox}
\usepackage{prettyref}
\usepackage{listings}
\usepackage{graphicx}
\usepackage{mathtools}
\usepackage{syntax}
\newrefformat{thm}{Theorem~\ref{#1}}
\newrefformat{cor}{Corollary~\ref{#1}}
\newrefformat{lem}{Lemma~\ref{#1}}
\newrefformat{cha}{Chapter~\ref{#1}}
\newrefformat{sec}{Section~\ref{#1}}
\newrefformat{app}{Appendix~\ref{#1}}
\newrefformat{tab}{Table~\ref{#1}}
\newrefformat{fig}{Figure~\ref{#1}}
\newrefformat{alg}{Algorithm~\ref{#1}}
\newrefformat{exa}{Example~\ref{#1}}
\newrefformat{def}{Definition~\ref{#1}}
\newrefformat{li}{Line~\ref{#1}}
\newrefformat{eq}{Equation~\ref{#1}}
\newrefformat{exa}{Example~\ref{#1}}

\definecolor{eminence}{RGB}{108,48,130}
\definecolor{commentgreen}{RGB}{2,112,10}
\definecolor{weborange}{RGB}{255,165,0}
\definecolor{frenchplum}{RGB}{129,20,83}

\usepackage{listings}
\pgfplotsset{
    discard if not/.style 2 args={
        x filter/.code={
            \edef\tempa{\thisrow{#1}}
            \edef\tempb{}
            \forcsvlist{\listadd\tempb}{#2}
            \xifinlist{\tempa}{\tempb}{}{}
        }
    }
}
\newcolumntype{P}[1]{>{\centering\arraybackslash}p{#1}}
\pgfplotsset{compat=1.18}

\begin{document}

\title{Compiling Recurrences over Dense and Sparse Arrays}

\author{Shiv Sundram}
\email{shiv1@stanford.edu}
\affiliation{%
  \institution{Stanford University}
  \streetaddress{P.O. Box 1212}
  \city{Stanford}
  \state{California}
  \country{USA}
}

\author{Muhammad Usman Tariq}
\affiliation{%
  \institution{Stanford University}
  \city{Stanford}
  \country{USA}
}

\author{Fredrik Kjolstad}
\affiliation{%
  \institution{Stanford University}
  \city{Stanford}
  \country{USA}}
\email{kjostad@stanford.edu}

\renewcommand{\shortauthors}{Sundram, Tariq, \& Kjolstad}

\begin{abstract}

Recurrence equations lie at the heart of many computational paradigms including dynamic programming, graph analysis, and linear solvers. These equations are often expensive to compute and much work has gone into optimizing them for different situations. The set of recurrence implementations is a large design space across the set of all recurrences (e.g., the Viterbi and Floyd-Warshall algorithms), the choice of data structures (e.g., dense and sparse matrices), and the set of different loop orders. Optimized library implementations do not exist for most points in this design space, and developers must therefore often manually implement and optimize recurrences. We present a general framework for compiling recurrence equations into native code corresponding to any valid point in this general design space. In this framework, users specify a system of recurrences, the type of data structures for storing the input and outputs, and a set of scheduling primitives for optimization.  A greedy algorithm then takes this specification and lowers it into a native program that respects the dependencies inherent to the recurrence equation.  We describe the compiler transformations necessary to lower this high-level specification into native parallel code for either sparse and dense data structures and provide an algorithm for determining whether the recurrence system is solvable with the provided scheduling primitives. We evaluate the performance and correctness of the generated code on various computational tasks from domains including dense and sparse matrix solvers, dynamic programming, graph problems, and sparse tensor algebra. We demonstrate that generated code has competitive performance to handwritten implementations in libraries. 
\end{abstract}
\keywords{}

\maketitle

\section{Introduction}

Recurrences compute the next value in a sequence based on previous values in the same sequence. They are used to express dynamic programs, graph analysis, linear solvers, and matrix factorization. Algorithms as diverse as sequence alignment, Dijkstra's algorithm, a triangular solver, and even Cholesky factorization can all be expressed as recurrence equations.

Two simple examples of recurrences are the Fibonacci sequence $F$ where $F_i = F_{i-1}+F_{i-2}$ and a running sum $S$ of an array $A$ where $S_i = S_{i-1} + A_{i}$. In both examples, the computed values depend on previously computed values, and there is a limited set of ways that values can be correctly computed without recomputing overlapping subproblems. The Fibonacci sequence cannot be sequentially computed in $O(N)$ time by filling the array backwards from $N$ to $0$, as it would violate the dependency $F_i$ has on $F_{i-1}$ and $F_{i-2}$, both of which have not yet been calculated. 

\begin{table}
\footnotesize
    \caption{Features and properties of different recurrences. X denotes the recurrence contains the feature}
    \begin{tabular}{c|c|P{1.5cm}|c|c|c|P{1.3cm}}%
    \bfseries Algorithm & \bfseries Type & \bfseries Diff Loop Orderings & \bfseries Mult Eqns & \bfseries Sparsity & \bfseries Masks & \bfseries Timestep Vars
    \begin{filecontents*}{features.csv}
Algorithm,Type,orderings,eqns,sparsity,masks,tvars
Prefix Sum,dynamic,,,,,
Cholesky,direct solver,X,X,X,X,
Triangular Solve,direct solver,X,,X,X,
Fused Tri Solve,direct solver,X,X,X,X,
QR,direct solver,X,X,X,X,
LU,direct solver,X,X,X,X,
Floyd Warshall,dynamic/graph,X,,,,X
Viterbi,dynamic/graph,X,,X,,X
Needleman-Wunsch,dynamic,X,,,,
Gauss-Seidel,iterative solver,X,,,,X
SpMV,tensor algebra,X,,X,X,
SDDMM,tensor algebra,X,,X,X,
\end{filecontents*}
\csvreader[head to column names]{features.csv}{}
    {\\\hline\Algorithm & \Type & \orderings & \eqns &  \sparsity & \masks &	\tvars }
    \end{tabular}
\label{tab:features}
\end{table}


Several performant library kernels for performing these operations exist. However, the set of recurrence implementations is a large design space across the set of all recurrences (e.g., the Viterbi and Floyd-Warshall algorithms), the choice of data structures (e.g., dense and sparse matrices), and the set of different loop orders. When a library implementation does not exist, programmers must write and hand-optimize their own implementations. Programmers are then faced with the challenge of manually balancing performance, portability across data structures, and developer productivity when writing their own kernels.

Existing approaches towards creating more general recurrence libraries and compilers do not capture this full design space. Dedicated systems for solving general recurrence equations in a bottom-up fashion, like the Dyna language \citep{dyna} which is similar to logic programming languages \citep{prolog}, require dynamic runtime analysis to track output dependencies and determine the order in which to calculate outputs. These languages are thus not as efficient as handwritten code which use static loop schedules. Furthermore, these systems do not generalize across data structures or optimization decisions like changing loop orderings, the latter of which requires reasoning about the recurrence's dependencies to check if this transformation is valid. The polyhedral model can reason about dependencies, but it similarly has no notion of sparsity or data structures. Some frameworks can  explore the design space of particular sparse recurrence problems, like GraphBLAS \citep{graphblas} for graphs and Sympiler for certain sparse solvers \citep{cheshmi2017sympiler}. However, neither is able to reason about dependencies for even the individual recurrences they target, and they therefore cannot automatically make optimizations like loop reordering. Other domains have successfully captured large program design spaces with compilers and DSLs like Halide \citep{halide} for image processing and TACO \citep{taco} for tensor algebra and sparse array programming \citep{sparseArrayProgramming}. However, no such compiler exists for recurrence equations, which includes all the algorithms listed in Table~\ref{tab:features}.

We thus introduce a set of techniques and abstractions for generating imperative code from independent specifications of declarative recurrence equations, the desired data structures, the desired loop nest ordering, and other optimization decisions. Based on these specifications, a compiler can generate bespoke code for CPUs, thus acting as a common domain specific language or library for recurrence problems. The contributions of this paper can thus be summarized as:

\begin{enumerate}

    \item An \textbf{intermediate representation} for describing the structure and data dependencies of an imperative loop nest that calculates a recurrence's outputs 

    \item A \textbf{code-generation algorithm} for lowering a declarative program of multiple recurrences and a loop nest ordering into this IR, which can then be lowered to C

    \item \textbf{Techniques specifically for optimizing recurrence programs}, including techniques for enabling loop fusion, auto-parallelization, and reductions in memory usage, which are enabled by this IR

\end{enumerate}

To evaluate these ideas,  we implement them in a compiler called RECUMA (RECUrrence computation MAchine), which we use to generate code for several recurrence problems from dynamic programming, matrix solvers, and tensor algebra. We demonstrate the compiler's ability to handle different schedules (specifying loop nest ordering and other optimization decisions) and data 
structure descriptions, showing how these choices affect performance. We embed within RECUMA a set of optimization strategies that can be shared across different recurrence problems. We show that the generated code's performance is competitive with handwritten implementations from existing libraries, including CXSparse \citep{cxsparse, directMethods} for sparse matrix decompositions, Parasail \citep{parasail} for sequence-alignment, Boost \citep{boostgraph} for Floyd-Warshall, and PolyBench \citep{PolyBench} for Gauss-Seidel. As tensor algebra is a subset of recurrence equations that have no dependencies amongst outputs, we also show that tensor algebra kernels generated by RECUMA have performance competitive with those generated by TACO.

\section{Cholesky Decomposition Example}

The prefix sum and Fibonacci examples in Section 1 were simple recurrence equations. Table~\ref{tab:features} shows that many important recurrences have  more complex properties. To more accurately illustrate the requirements of a recurrence compiler and the complexity it must capture, we instead use the Cholesky decomposition as a running example. For any symmetric positive definite matrix A, the Cholesky decomposition computes a lower triangular matrix L such that $LL^T=A$. The decomposition is commonly used to solve the system $Ax=b$, in which the decomposition of $A$ is followed by two triangular solves to compute $x$. This decomposition is expressible as a system of two mutally dependent recurrences defined on the same output matrix L, where previously computed portions of L are used to compute subsequent entries:

\[ L_{ij} = \frac{A_{ij} - \sum_{k=0}^{j} L_{ik} L_{jk}}{L_{jj}}  \;\; : \;\; j<i \; \; \; \; \; \; \; \hspace{0.5cm} 
L_{ij} = \sqrt{ A_{ij} - \sum_{k=0}^{i} L_{ik} L_{jk}} \;\; : \;\; i=j \] 

Each recurrence contains mathematical operations over a set of 2D indexed tensors. Each recurrence also has an independent set of constraints that bound the index variables' iteration spaces. In this case, the constraints demonstrate how the first recurrence calculates lower triangular elements of the output ( where $j<i$), while the second recurrence calculates the diagonal entries (where $i=j$). There are implicit dependencies within and between the equations. The calculation of the diagonal entries relies on non-diagonal values of $L$, and vice versa. Any valid Cholesky decomposition program must respect these intra and inter-recurrence dependencies.

\begin{figure}
    \centering
    \includegraphics[scale=.5]{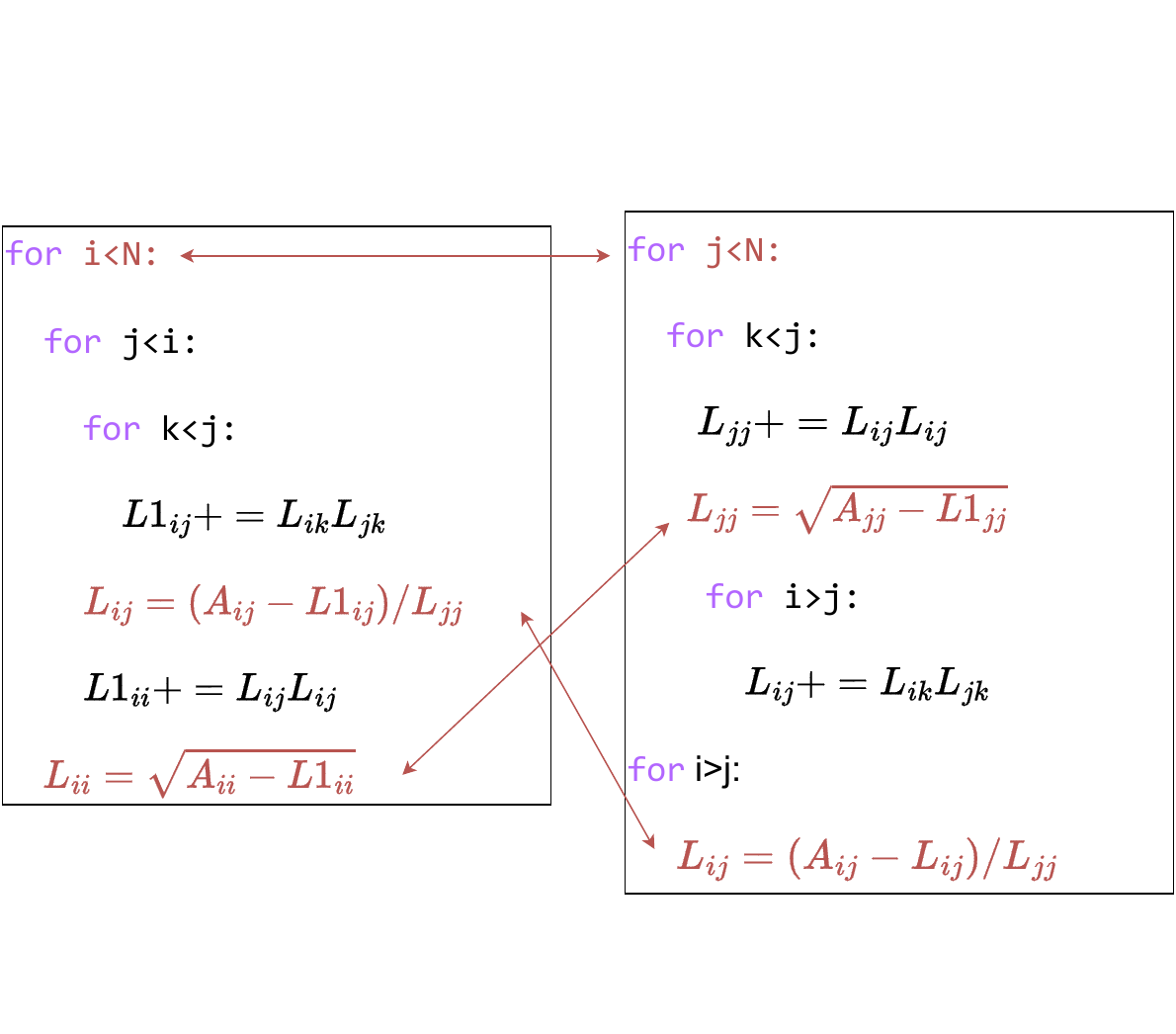}
    \caption{$ijk$ vs $jki$ Cholesky. Several differences between corresponding recurrence computations are highlighted in brown and marked with arrows. The relative placement/order of certain computations are necessarily permuted in the two versions. }
    \label{fig:compPseudo}
\end{figure}

\begin{figure}
    \centering
    \begin{lstlisting}[language=C++]
for(int i=0; i<N; i+=1) {
    //not shown: load sparse row i of A into dense array tmpAij
    //init dense workspay array tmpLij to build output row i of L
    double Lii = 0;
    for(int j=0; j<i; j+=1) {
        double L1ij = 0;
        for (int p_k_Ljk = L1_pos[j]; p_k_Ljk< L1_pos[j+1]; p_k_Ljk++) {
            double Ljk = L_vals[p_k_Ljk];
            double Lik=tmpLij_vals[Ljk_crd];
            L1ij += Lik * Ljk;
        }
        double Aij = tmpAij_vals[j];
        double Ljj=L_vals[L1_pos[j]-1];
        double Lij = (Aij - L1ij) / Ljj;
        tmpLij_vals[j] = Lij;
        Lii+= Lij*Lij;
    }
    double Aii=tmpAij_vals[i];
    tmpLij_vals[i] = Lii = sqrt(Aii-Lii);
    //not shown: compress tmpLij into new row of output L
    //set L_1crd and L1_pos
}
    \end{lstlisting}
    \vspace{-0.8em}
    \caption{Sparse $ijk$ cholesky}
    \label{fig:cholSparse}
\end{figure}


Implementing a Cholesky decomposition with different loop orderings requires significantly different programs. This is in contrast to other matrix kernels (e.g. matrix-matrix multiply, matrix-vector multiply), in which permuting the respective loop headers is all that is required to implement different loop orderings. To illustrate, pseudocode for the commonly-used "up-looking" Cholesky decomposition is shown on the left side of Figure~\ref{fig:compPseudo}. We denote the loop nest as an $ijk$ form of the Cholesky decomposition given the nesting order of the loops and each loop's iteration variable. This $ijk$ loop ordering, otherwise known as the Cholesky–Banachiewicz algorithm, is useful when the underlying matrices $A$ and $L$ are stored in a row-major fashion, which will ensure good memory access locality. If $A$ and $L$ are column-major, however, it is better to use a loop nest with a $jki$ loop ordering, shown on the right side of Figure~\ref{fig:compPseudo}.

Besides the aforementioned difference in the loop nesting order, we note several differences between the $jki$ and $ijk$ formulations:

\begin{description}
    \item[Loop Headers] The loop headers corresponding to the loops over $j$ and $i$ are different due to the recurrence constraints. In the $ijk$ loop nest, $i$ iterates from $0$ to $N$, whereas in the $jki$ loop nest it iterates from $j+1$ to $N$. 
    \item[Diagonal Elements ] The calculation of the diagonal elements $L_{ii}$ in the $ijk$ loop nest are reformulated as calculations of $L_{jj}$ in the $jik$ nest.
    \item[Number of Loops ]The two implementations have a different number of for loops. The $jki$ loop nest has an additional loop over $i$ at the end of the computation to scale non-diagonal element $L_{ij}$ by the diagonal $L_{jj}$. Unlike the $ijk$ nest, this additional loop cannot be fused into the previous loop over i, as $L_{jj}$ is not fully calculated until the first loop over $i$ completes.

\end{description}

Therefore, unlike a matrix multiply kernel where an $ijk$ loop nest can be converted into a $jki$ loop nest simply by permuting the order of the three nested loop headers, switching the loop iteration order in a recurrence system may require more changes. We note the existence of six possible Cholesky loop orderings corresponding to each permutation of $ijk$ ($ijk$, $kji$, etc.), all of which have notable differences from each other and are useful in different scenarios. Despite these differences, all six variants share that common recurrence equation formulation. Similar recurrences systems can be defined for the LU decomposition, QR decomposition, and triangular solve. 

Finally, when a large portion of the entries of $A$ are zero, we may want to generate an implementation of any of the six Cholesky loop orderings that uses sparse data structures. An example is seen in Figure~\ref{fig:cholSparse} for an $ijk$ loop ordering. This code is more complicated than the dense version, but we would still like to compile the recurrence into either dense or sparse code based on user description of $A$'s and $L$'s data structures.

\section{Overview}

Our recurrence compiler takes in a declarative system of recurrences, dense/sparse data structure descriptions, a loop ordering, and optional optimization options. The latter two are together referred to as the schedule. The compiler generates an imperative C program that solves the desired equations. The generated code calculates the provided recurrences' outputs while respecting dependencies, adhering to the loop ordering, and targeting the user-specified data structures. 

\begin{figure}
    \centering
    \includegraphics [scale=.6]{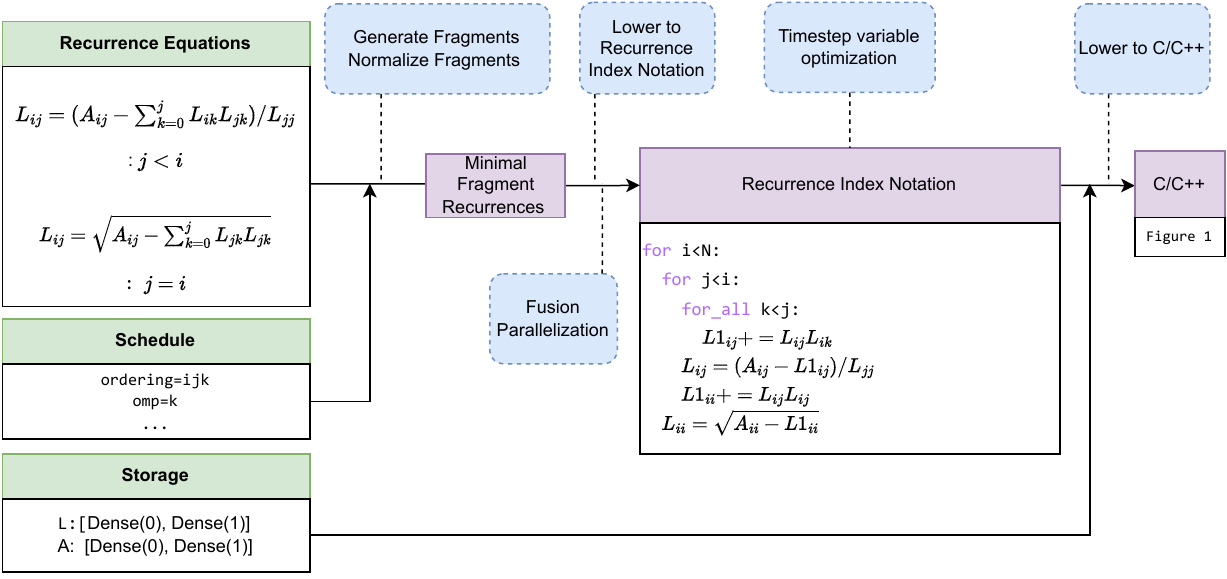}
    \caption{An overview of the recurrence compilation workflow used by RECUMA. Green boxes  represent user-provided inputs to the compiler. Purple boxes are intermediate or final outputs of the compiler's lowering passes. Blue boxes represent passes of the compiler, including those for lowering and optimization.}
    \label{fig:Overview}
\end{figure}

Moving from declarative expressions of mathematical recurrences to an imperative C program requires a series of lowering passes shown in Figure ~\ref{fig:Overview}, which provides an overview of a recurrence compiler workflow. These passes operate on the mathematical recurrences equations themselves and then on an imperative intermediate representation we call Recurrence Index Notation. Figure ~\ref{fig:Overview} illustrates this process for a dense Cholesky composition adhering to an $ijk$ loop ordering, ultimately resulting in the C program shown on the left side of Figure~\ref{fig:compPseudo}.


The first pass decomposes each recurrence into its minimal fragment recurrences. This decomposition converts each original recurrence, whose right-hand side expression may include many mathematical operations, into multiple new recurrences that separate summation and reduction operations from operations working on the reduced value. This decomposition process is a preparation stage for the subsequent lowering algorithm, which takes a set of recurrences as an input and then places each recurrence's right-hand side expression into a location in the loop nest. This decomposition allows the lowering algorithm to work at the granularity of placing subexpressions from the input recurrences. 

The minimal fragment recurrences are then lowered into the Recurrence Index Notation (RIN) intermediate representation. RIN is an imperative IR in which all data accesses are logical, allowing the compiler to reason about the desired program's loops, computations, and dependencies without regard for how the tensors are stored. Therefore, loop reordering, dependency analysis, placement of compute statements within loop nests, and low-level program optimizations (e.g. loop fusion and auto-parallelization) can be carried out without the need to manage low-level constructs related to sparse code, such as while loops, if statements, and indirect memory accesses.

 The minimal fragment recurrences are lowered to RIN with a new code-generation algorithm. This algorithm generates an RIN representation describing the program's loop structure and placement of compute statements. In this process, the subexpressions associated with each minimal fragment are greedily placed into the first location of a loop nest where the expression's dependencies are considered satisfied. The core component of this algorithm is its ability to inductively predetermine about which portions of the output can be considered already calculated at each iteration of the loop nest.  When it is not possible to generate a program that respects the desired loop ordering or the dependencies of the recurrences, the compiler reports an error, denoting that the provided combination of recurrences and loop ordering is not possible. 

Once the RIN is fully formed, it is possible to apply user specified optimizations to it. Loop fusion happens automatically in the RIN generation stage, but other optimizations can be done by inspecting or modifying the RIN. The final pass lowers RIN into C code. We use standard techniques utilized by tensor algebra compilers like TACO, COMET \cite{comet}, and MLIR-sparse \citep{mlirsparse} for generating sparse loops, augmented to support dependent loop bounds. The final lowering pass is the only pass that requires information about the data structures in which the tensors are stored, which by design was made irrelevant to the preceding compiler passes. The final code can  use dense arrays or sparse data structures, which can be used to represent graph data structures like adjacency matrices and adjacency lists.

\section{Recurrence Language}

The abstract language of potential inputs to our compiler is based on mathematical notation commonly used to define recurrence equations. The language consists of a set of equations for computing scalar arithmetic expressions over tensors. These tensors are indexed by index variables, and scalar expressions can be summed over reduction indices. The language is thus similar to that of the tensor index notation \citep{ricciTensorIndex}, a language of binary operations, unary operations, and contractions over a set of indexed tensors that is used by TACO. However, our language of recurrences contains three properties that are not supported in the tensor index notation:

\begin{description}
    \item[Dependencies] The core properties of a recurrence equation are the dependencies between computations of output elements, as seen in all the aforementioned recurrence equations. The same tensor may appear on both the left and right-hand sides of a recurrence equation. 
    \item[Constraints ]  A constraint bounds the iteration space of one index variable in terms of another index variable. These constraints   are either equalities or inequalities between index variables (e.g. $j<i$, $j=i$). 
    \item[Multiple equations] A recurrence problem may consist of multiple individual recurrence equations. The equations can use the same output as their left-hand sides, as long as each recurrence computes a disjoint part of the output. Furthermore, these recurrences can be mutually dependent; output elements computed in one recurrence equation may be used to calculate outputs of a 2nd recurrence, and vice-versa. 
\end{description}

All three properties are notably present in the Cholesky decomposition, and the aforementioned equations for the Cholesky decomposition are expressed in this language. Each program in this language is thus a set of recurrence equations, in which each equation has its own set of constraints over index variables. Constraints may be implicitly defined by the bounds of a summation used within the recurrence (e.g. $k<i$ in the Cholesky equations) or defined separately and applied to the recurrence (e.g. $j<i$ in the Cholesky equatations).



\begin{figure}[tb]
\AtBeginEnvironment{grammar}{\scriptsize}
\newenvironment{BNF}
  {}

\begin{BNF}
\label{grammar:my-grammar}
\begin{multicols}{2}
\begin{grammar}
<iVar> ::= string 

<index> :: = iVar `+' ConstantInt
\alt            ConstantInt


<Tensor> ::= string


<TensorAccess> ::= $Tensor_{index^+} $

<expr> ::= BinaryOp`( 'expr, expr `)'
 \alt            UnaryOp`(' expr `)'
 \alt            TensorAccess 
  \alt           $\sum_{iVar=index}^{index}$ expr
 \alt            Constant

\columnbreak

<Recurrence> ::= TensorAccess `=' expr

<Constraint> ::=  iVar `<' index
\alt 			iVar`<=' index
\alt			'iVar `=' index

<Constraints> :: = `[' Constraint (`,' Constraint )* `]' 

<ConRecurrence> ::= Recurrence `:' Constraints 

<Program> :: = `[' ConRecurrence (`,' ConRecurrence )* `]' 
\end{grammar}
\end{multicols}
\vspace{-0.8em}
\caption{Grammar for the recurrence language
\label{fig:gram1}
}
\end{BNF}
\end{figure}

\subsection{Expressing Recurrence Equations}

The grammar of this recurrence language is presented in Figure~\ref{fig:gram1}. A $\langle \text{Recurrence} \rangle$ represents a simple recurrence equation connecting an indexed element of a tensor and the expression required to compute it. Each tensor element is a $\langle \text{TensorAccess} \rangle$, which denotes an  element indexed by a sequence of $\langle \text{index} \rangle$s, which each represents an index variable optionally offset by a constant. A single TensorAccess is used to form the left-handside of a Recurrence. A Recurrence with an optional list of constraints is a $\langle \text{ConRecurrence} \rangle$, a set of which makes a program.

A TensorAccess's indexing expression is constrained such that the expression used to access a specific dimension of a tensor consists of a single index variable added to an optional integer constant. This is illustrated by the Fibonacci equation $F_i = F_{i-1} + F_{i-2}$, where the indexing expressions on the right-hand side are $i-1$ and $i-2$. While this set of expressions does not include the full spectrum of affine indexing expressions, it includes those used by a large subset of recurrences. In theory, the constraint need not be this restrictive; in principle it is only required that for a pair of TensorAccesses $F_{f(i)}$ and $F_{g(i)}$, where $f$ and $g$ are functions over $i$, that a program/compiler can mathematically determine whether $f(i)>g(i)$ for all $i>=0$. The constraint of having $f(i)$ and $g(i)$ each be a single index variable with a constant offset makes this determination trivial. This requirement is used by our greedy lowering algorithm when determining which parts of the output can be considered fully calculated at each iteration of the loop.

A Recurrence equation's right-hand side is formed of mathematical operations and reductions which operate on one or more TensorAccesses and Constants. The list of currently supported BinaryOperations include addition, subtration, multiplication, min, max, and can easily be expanded. Similarly, UnaryOps include operations like square roots and negations. Each recurrence equation is assigned a unique set of constraints defined over index variables.

The language includes an operator denoting a reduction/summation across an index variable (iterating from an initial value towards an upper bound) for a given expression. This reduction can be built on top of any commutative and associative binary operation. The reduction statement embeds a constraint between the summation variable and this upper bound. As such, this constraint does not need to be described separately. 

\subsection{Denoting Schedule and Storage }

The main component of a schedule is an ordered list of index variables denoting loop ordering. As shown, a Cholesky decomposition can be implemented as a triply nested loop. If the user wants to generate a program in which the outer loop iterates over $j$, the middle over $k$, and the inner over $i$, then the user will simply provide a $jki$ loop ordering to the compiler. 

The final component is the description of the input and output data’s storage format. Each input and output tensor can have multiple dimensions, and each dimension must have a level format (Dense or Compressed). The Dense keyword states that in a particular dimension, each element is stored in the data structure, including zero entries. The Compressed keyword denotes that for a particular dimension, only nonzero elements are stored. These level formats can be composed \citep{taco} to form common sparse formats like Compressed-Sparse Rows (CSR) or Compressed-Sparse Columns (CSC). To generate a Cholesky decomposition over a CSR matrix A, the matrix’s associated storage type is  [Dense(0), Compressed(1)]. The 0 before 1 indicates that rows are stored before columns. Accordingly, in a column-major matrix dimension 1 is stored before 0. 

By separating specifications of equations, constraints, schedules, and storage info, the user can then concisely express a variety of recurrences over dense tensors, sparse tensors, and graphs, which are ultimately either dense or sparse matrices. For a Cholesky decomposition using CSR matrices, the recurrences, storage, and loop ordering are expressed to RECUMA as so:

\begin{lstlisting}[language=Python,numbers=none,frame=none]
rec1 = "L(i,j) = (A(i,j)-Sum{k}(L(i,k)*L(j,k)))/L(j,j) : [k<j,j<i]"
rec2 = "L(i,j) = sqrt(A(i,j)-Sum{k}(L(i,k)*L(j,k))) : [k<j,j=i]"
schedule["ordering"] = "ijk"
storage["A"] = Storage([Dense(0), Compressed(1)])
storage["L"] = Storage([Dense(0), Compressed(1)])
program = Program([rec1, rec2], schedule, storage) #generate C program
\end{lstlisting}

\section{Recurrence Dependencies}
Recurrence computations contain dependencies between outputs, meaning the computation of outputs may not be embarrassingly parallel. These outputs, therefore, cannot be calculated in an arbitrary order. When the user provides a loop ordering, it imposes a constraint on the order in which to calculate outputs. By identifying these dependencies in the recurrences equations, representing the dependencies in a DAG, and then reasoning about how loop ordering affects dependency management, we can construct the foundation of an algorithm for lowering recurrences to imperative code. 

\begin{figure}
    \centering
   \includegraphics[scale=.37]{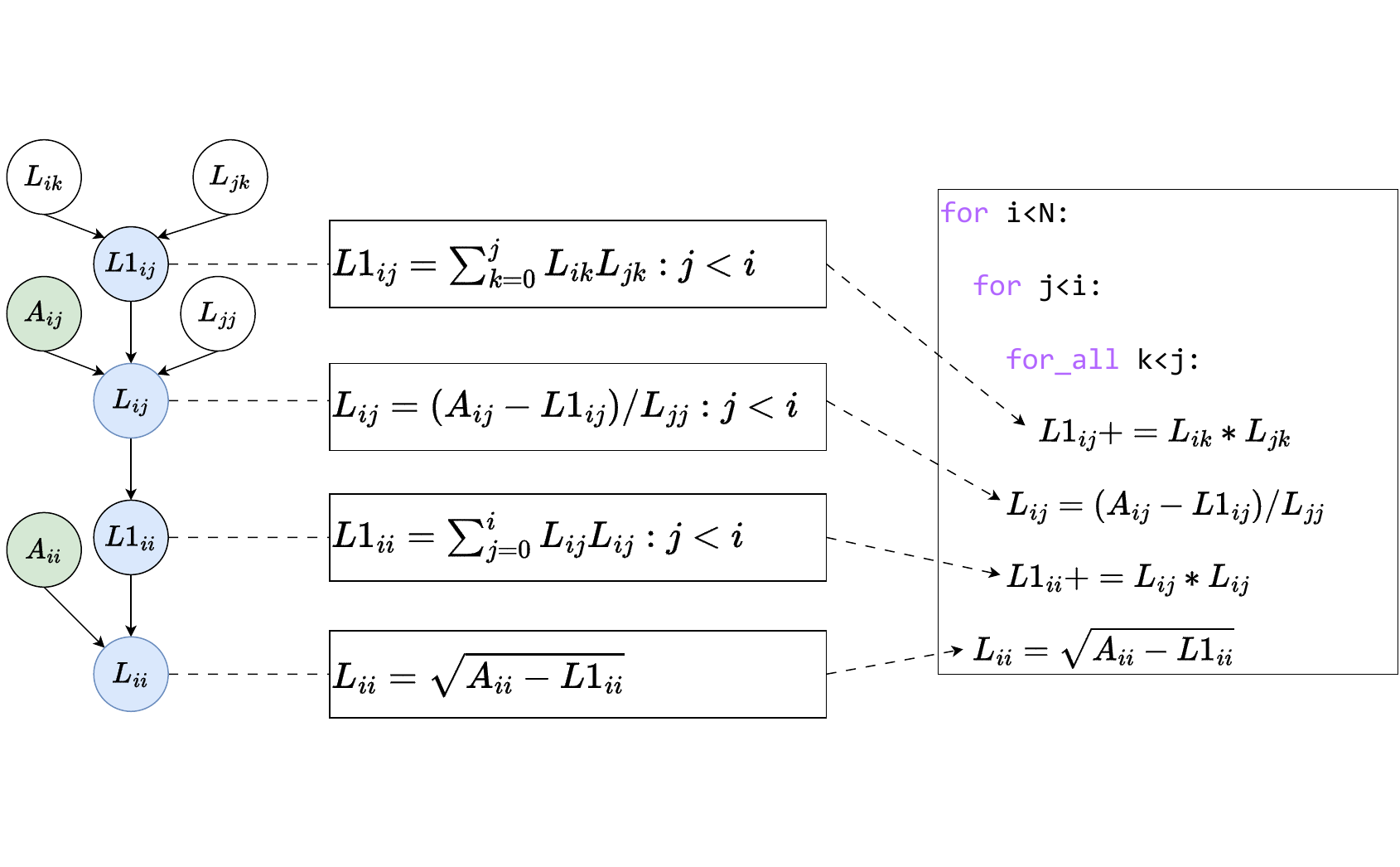}
    \caption{Fragment recurrences, corresponding DAG nodes, and corresponding placements for $ijk$ Cholesky}
    \label{fig:deps}
\end{figure}

\subsection{Dependency Graphs}
In a recurrence, a \emph{dependency} is any previous element in the recurrent sequence used to calculated the current element. A recurrence equation lists its output on its left-hand side and its operands on the right-hand side. All dependencies can therefore be identified by inspecting the input recurrence equations, in which a dependency exists between a recurrence's output term and each of the right-hand side terms. All of these input and output terms correspond to a TensorAccesses. In our Cholesky example, the output term is $L_{ij}$, and the dependencies include $L_{ik}$, $L_{jk}$, $L_{jj}$, and $A_{ij}$. We model these dependencies with a DAG, in which each TensorAccess (representing a term in the recurrence) is a node and each dependency is an edge. 

Before identifying dependencies in a recurrence system, recurrences are broken into smaller, \emph{minimal fragment recurrences}. A minimal fragment recurrence is a recurrence in which any summation/reduction statement is placed in a separate equation from any operations on the summed/reduced value. In our  Cholesky decomposition example, the two input recurrences are accordingly reduced to the four minimal fragment recurrences shown in the center column of Figure~\ref{fig:deps}.

Any recurrence can be decomposed into a set of minimal fragments, which each represent a subexpression from the original recurrence. The subsequent code generation stage places the minimal recurrences into the loop instead of the original recurrences. Decomposing a recurrence into fragments allows subexpressions from different input recurrences to be interleaved and placed together into a single fused loop nest. 

For each \emph{minimal fragment recurrence}, an edge is drawn from each right-hand side operand to the corresponding output term on the left left-hand side. Figure ~\ref{fig:deps} shows how terms in the minimal fragment recurrences are translated into nodes and edges in this DAG. The subsequent lowering algorithm then compiles this DAG into an imperative RIN, shown on the right side of Figure~\ref{fig:deps}.

\begin{figure}
    \centering
   \includegraphics[scale=.47]{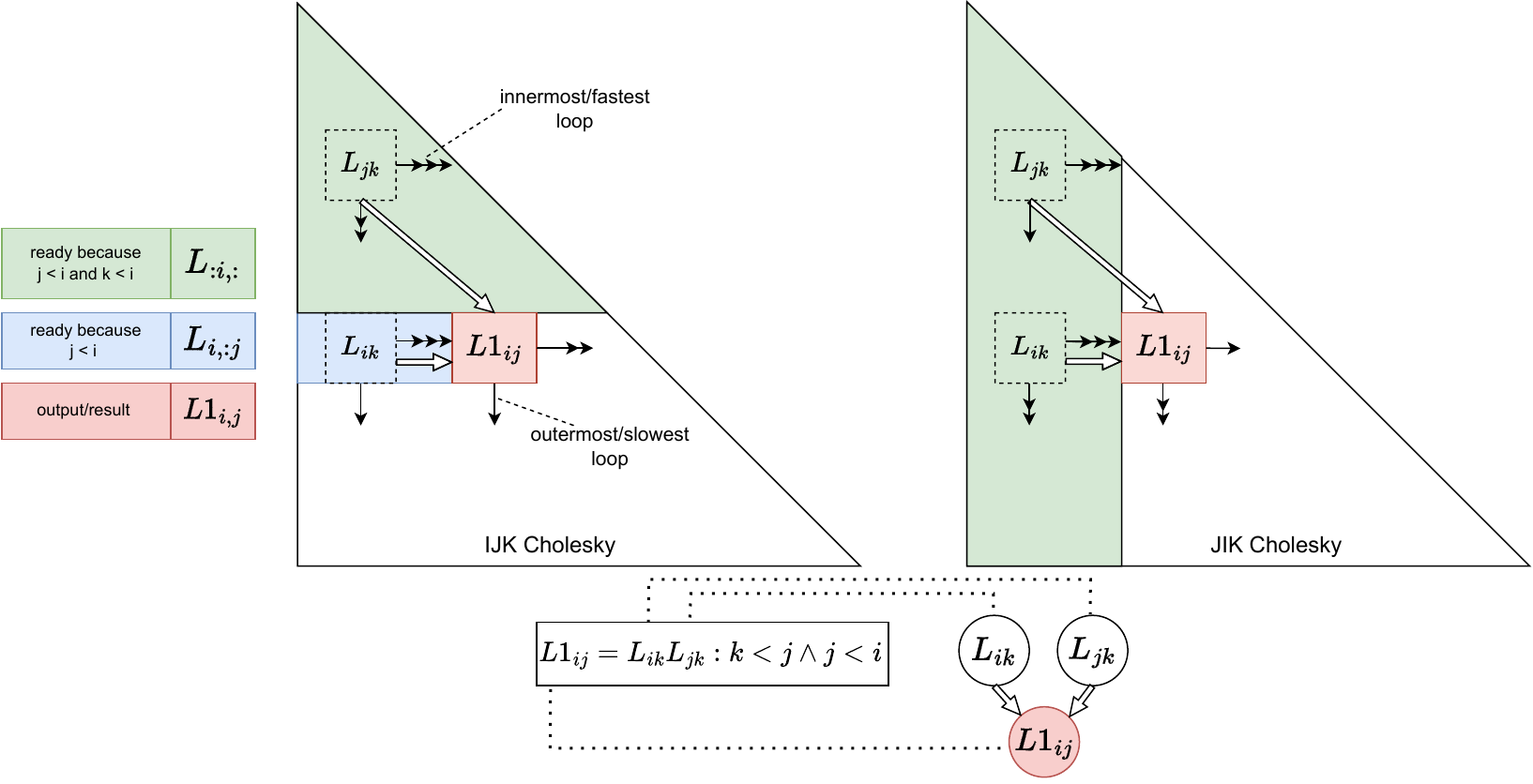}
    \caption{Comparing dataflow and dependencies of $ijk$ and $jik$ Cholesky decompositions. Calculating $L1_{ij}$ requires that $L_{ik}$ and $L_{jk}$ are calculated. Solid arrows show the directions of data iteration in the  triply nested Cholesky loop. For an $ijk$ Cholesky program, which calculates the output one row at a time, two successive inductive assumptions (green and blue) are required to satisfy the dependencies. For a $jik$ Cholesky program, which calculates the output one column at a time, only one inductive assumption is necessary.}
    \label{fig:compareCholDiagram}
\end{figure}

\subsection{Managing Dependencies}
Any code generation algorithm for recurrences must manage recurrence dependencies such that the final code does not violate them. In particular, the treatment of dependencies depends both on the recurrence and the user-requested loop ordering. For many recurrences, certain loop orderings are actually impossible, and for others, certain loop orderings will have different loop-carried dependencies. 

The affect of different loop-orderings on dependency handling is exemplified by comparing different loop orderings of the Cholesky example. Figure~\ref{fig:compareCholDiagram} shows the dependencies necessary to calculate $L1_{ij}$ for two different loop orderings of a Cholesky decomposition. The right side shows a $jik$ schedule in which the output $L$ is calculated column-by-column. The dependency DAG shows that calculating $L1_{ij}$, an element in column $j$, requires having already calculated the two dependencies $L_{ik}$ and $L_{jk}$. In this loop ordering, we calculate the output column-by column, such that the $j$th iteration of the outer loop fully computes column j. Accordingly, the Cholesky constraints state that $k<j$, so we know that column $k$, which includes both dependencies, is already calculated by the time we get to calculating column $j$.

Meanwhile, computing $L$ with $ijk$ schedule, as shown on the left side of Figure~\ref{fig:compareCholDiagram}, will calculate $L$ row-by-row. In this case, the dependencies of $L1_{ij}$ are still $L_{ik}$ and $L_{jk}$, which are in rows $i$ and $k$ respectively. However, when  calculating $L1_{ij}$ in row $i$, the constraints $k<j<i$ only guarantee that $L_{jk}$, in row $j$, is already calculated. The other dependency, $L_{ik}$, is not satisfied, as it is in the \emph{same} row $i$ we are trying to calculate. 

However, we can satisfy the dependency $L_{ik}$ by assuming that when calculating row $i$, that we calculate elements \emph{within} the row in a Cartesian order, starting from column zero ($L_{i,0}$) and until the last column $i$ ($L_{ii}$). In this case, because $j<i$, we know that the $j$th element in the current row will already be calculated by the time we are calculating the $i$th element.

These properties require that in the $jik$ program, the outer loop over $j$ will contain a loop carried dependency. This reflects that fact that when calculating an entry in column $j$, we must have already calculated all previous columns until column $j$. Meanwhile, in the $ijk$ program, \emph{}{both} the outer loop over $i$ and middle loop over $j$ will contain loop carried dependencies; when calculating an entry in row $i$  we must have calculated all previous rows up to $i$. Additionally, within a row, when calculating the $j$th entry $L1_{ij}$ we must have already calculated all previous entries in the row until the $j$th one. This additional constraint results in the middle loop over $j$ containing a loop-carried dependency. 

A loop with a loop carried dependency may not be trivially parallelized, unless the dependency emerges from a reduction operation. In $ijk$ Choleksy, because there is a loop carried dependency in the outer and middle loop, only the inner-most loop can be parallelized. Meanwhile, in  $jik$ Cholesky, only the outer loop contains a loop carried dependency, meaning both the middle and inner loop can both be parallelized. 

Our RIN generation algorithm manages recurrence dependencies with \emph{inductive assumptions} that formalize the above intuition A proof by induction makes an assumption that a statement over $j$ is true for all $j<i$, and then uses this assumption to prove that the same statement is true for $i$. If both this inductive proof and the base case are true, then the statement is always true. Similarly, if we can calculate the $ith$ row of $L$ by assuming that all rows $j<i$ are already calculated, then we can always satisfy the dependencies necessary to calculate every entry in $L$.

\section{Recurrence Index Notation Code Generation}
Code generation lowers recurrences into the imperative Recurrence Index Notation (RIN) IR. This IR, described in Section~\ref{sec:RIN}, represents a program in which the recurrences' outputs are iteratively calculated in an abstract loop nest, where physical data structures have not yet been specified, whose nesting order matches the user's specifications. The process of lowering recurrences into the RIN is based on three principles, each of which forms a step in this algorithm. 
\begin{enumerate}
    \item Recurrences can be decomposed into minimal fragments that each represents a single assignment statement with one or two operands.
    \item RIN can then be generated by placing the minimal fragments into a loop nest one at a time, such that dependencies between fragments are not violated. Dependencies can be modeled with a DAG that also encodes the order in which to place statements.
    \item When determining where to place a statement into the loop nest, it is possible to inductively determine which outputs are already calculated at any location and any iteration of the nested loop. This is necessary when the statement depends on outputs computed by this same statement but in earlier iterations .
\end{enumerate}
This code generation algorithm follows these three steps to form an imperative program in the RIN and terminates once all fragments have been placed. Pseudocode for the full placement algorithm is shown in Figure~\ref{fig:pseudo} and described in the following sections.



\begin{figure}
\begin{minipage}{\linewidth}
\begin{lstlisting}[language=Python]
#1) generates minimal fragements 2) sorts fragments 3) places each fragment into RIN program, making inductive assumptions when necessary
def lower(input_recurrencs, loop_ordering):
    fragment_recurrences = generateFragments(input_recurrences)
    sorted_fragments = topolocialSort(fragment_recurrences)
    assumes = {} #no inductive assumptions yet
    rin_program = {} #RIN program starts as empty 

    for statement in sorted_fragments:
        tryPlacing(statement, rin_program)

        #try speculating inductively and inserting extra readiness markes
        while statement.notPlaced() and canMakeAssumption(loop_ordering, assumes):
            assumes = makeAssumption(loop_ordering, assumes) 
            tryPlacing(statement, rin_program, assumes) 
            
        if statement.notPlaced(): 
        return fail #loop_ordering is impossible
    return rin_program

#greedily place statement into first location where dependencies are satistfied
def tryPlacing(statement, program, assumptions):
    #function defined by Equation 2
    location = PlacementLocation(statement, statement.dependencies, rin_program)
    if location:
        location.place(statement) #inserts readiness statement
\end{lstlisting}
\end{minipage} 
\caption{Pseudocode for RIN generation. Not shown is 1. the process for generating empty loop headers as a potential location for placing a fragment and 2. the check that index variables used in a fragment are in scope at the potential location. Both of these happen implicitly when iterating over valid locations in the program}
\label{fig:pseudo}
\end{figure}

\subsection{Recurrence Index Notation}
\label{sec:RIN}
The code generation process first  generates Recurrence Index Notation (RIN), an imperative intermediate representation of the program. The RIN consists of abstract loops that let us reason about program dependencies, loop nest order, and the placement of compute statements without regard to the program's data structures and how they are iterated over. Hence, RIN contains no notion of the underlying data structure being used and all accesses to a tensor are logical. The process of placing compute statements into a loop nest for a specific loop ordering while respecting dependencies is thus done at the RIN level. 

In RIN, the ordering of loops is specified by the user, and the iteration bounds of each loop header is determined by how the iteration variable $u$ is bounded in the constraints, filtered by which variables are already in the present scope of the loop header. That is, variable $v$ can only form the lower or upper iteration bound for a loop over $u$ if this loop is already nested within a loop over $v$. For example, given a $jik$ loop ordering and the constraints $0 \leq k<j<i<N$, $j$'s bounds will be $O<=j<N$, $i$'s bounds will be $j<i<N$, and $k$'s bounds will be $0=<k<j$. 

An example RIN loop nest is shown in Figure~\ref{fig:Overview}. The grammar for RIN is shown in Figure~\ref{fig:RINgram}. The if statements and while loops often found in sparse codes are absent in RIN; all loops are for or forall loops (in which a for loop has loop-carried dependencies, and a forall does not). Therefore, a recurrence program that requests dense data structures and an otherwise equivalent recurrence program that requests sparse data structures will have the exact same RIN. 

RIN loop nests also contain non-compute readiness markers that track what outputs are already computed at different points in the loop. Each readiness marker indicates that a certain subsection of the output has been computed at any point after the statement's location in the RIN. These readiness markers are placed into RIN whenever fragments are placed and documents inductive assumptions. In the Cholesky example, for instance, a readiness marker denoting that $L_{ii}$ is calculated by the statement's location would be "$//ready\; L(i,i)$."

\begin{figure}[tb]

\AtBeginEnvironment{grammar}{\scriptsize}

\begin{grammar}
\begin{multicols}{2}
<Tensor> ::= string

<assign> ::= TensorAccess `=' expr
 \alt       TensorAccess `+=' expr

<expr> ::= BinaryOp`( 'TensorAccess, TensorAccess `)'
 \alt      UnaryOp`(' TensorAccess `)'

<index> ::= iVar 
\alt        iVar `+' ConstantInt
\alt        ConstantInt

<rIndex> ::= index
\alt            `:'
\alt            `:' index

<readiness> :: = \newline
`//ready' Tensor `(' rIndex (`,' rIndex )* `)' 

<TensorAccess> ::= Tensor `(' index (`,' index )* `)' 

<stmt> ::= assign
\alt        for
\alt        forall
\alt        readiness

<stmtlist> ::= stmt+

<loopheader> ::=  iVar `<' index  `:'
\alt			 index `<' iVar `<' index `:'

<for> ::= `for' loopheader `:' stmtlist

<forall> ::= `forall' loopheader `:' stmtlist

<RIN> ::= stmtlist

\end{multicols}
\end{grammar}
\vspace{-1.0em}
\caption{Grammar for Reccurence Index Notation}
\label{fig:RINgram}
\end{figure}

Let $R$ be a collection of locations in a RIN loop nest, ordered according to program order. At any particular location $p \in R$, the set of outputs and temporaries that have already been computed at $p$ can be calculated with the following equation:

\begin{equation} \label{eq:ready}
readyAtLocation(p) =  \{out \; | \; ready(out) \text{ in lexical scope of } p \}
\end{equation}
Therefore, the set of already calculated outputs at point $p$ is the set union of all readiness markers in program locations before $p$. In other words, it is determined by an in-order traversal over all readiness markers in the RIN's AST, terminating at location $p$.

\subsection{Generating Minimal Fragments}
Given a recurrence, its set of minimal fragments is generated by traversing the recurrence bottom-up and placing each summation operation in its own recurrence. Because these new fragment recurrences do not compute a final output, they instead store their partial sums to temporaries. To disambiguate a temporary result from a result, the temporary result is denoted by the corresponding output tensor name $A$ being followed by a number (e.g. $A0$, $A1$). The temporary tensors are named aliases of the final output tensor, such that each alias contains the partial computation of a value that will eventually reside in the output tensor. The Cholesky decomposition, for example, will decompose each of its two input equations into two minimal fragments, since each equation contains one summation. The four minimal fragment recurrences for the Cholesky decompositions are shown in Figure~\ref{fig:deps}.


\subsection{Determining Placement Ordering}
The placement algorithm only places a fragment into the loop at a location where the fragment's dependencies are considered already calculated. There is thus a limited set of correct orderings in which different fragments should be placed into the imperative loop, and these orderings can be inferred from a DAG over all fragments' TensorAccesses. This ordering is determined after fragment generation and before the placement stage.

In a minimal fragment graph, a directed edge indicating a dependency exists from TensorAccess $B$ to TensorAccess $A$ if there exists a minimal fragment with $A$ on the left-hand side and $B$ on the right-hand-side (i.e. if $B$ is required to calculate $A$). The set of TensorAccesses across all fragments forms a directed acyclic graph, in which each TensorAccess is a node and each dependency is a directed edge. The placement order is then determined by a topological sort, which occurs directly after generation of minimal fragments. For example, in an $ijk$ Cholesky decomposition, the first placed fragment is $L1_{ij} \mathrel{+}= L_{ik}L_{jk}$ because of its position in Figure~\ref{fig:deps}; of the four nodes referring to fragments' outputs, it is the only one in the DAG not dependent on the other three. $L1_{ii}$, on the other hand, cannot be placed first. As shown in Figure~\ref{fig:deps}, it is dependent $L_{ij}$, which itself is dependent on $L1_{ij}$, the aforementioned output node. 


\subsection{Greedily Placing Minimal Fragments into a RIN Loop Nest}
\label{sec:greedy}

The placement algorithm starts with an empty loop nest created from the recurrence's index variables and a user-specified loop order. The algorithm proceeds by iterating through the minimal fragment DAG in topological order and placing each fragment into a location within the RIN. 

For a statement $S$ with a set of dependencies $D$ (the fragment's parent nodes in the DAG) the placement algorithm chooses a location $l$ in RIN $R$ in which to place $S$. Assuming locations in $R$ are ordered according to an in-order traversal of the RIN's AST, this location can be determined using Equation~\ref{eq:ready} calculating $readyAtLocation(p)$. 

\begin{equation} 
placementLocation(S,D,R) = \min_{\forall p \in R} p \; \; \; s.t. \;\; D \subset readyAtLocation(p) 
\end{equation}
In other words, each statement is greedily placed into the first location in the RIN loop nest where the statement's dependencies are computed.

When a statement is placed in the loop nest, a new readiness marker gets placed directly after it. This marker indicates that the left-hand side calculated by the statement has now been computed. This placement algorithm terminates when all fragments have been placed, in which case the IR is fully formed, or when it it not possible to place a statement anywhere. In this last case, the given loop-ordering is illegal and an error is thrown.

\subsubsection{Inductive Assumptions}
A statement may depend on a value in the output calculated by itself in an earlier iteration of the loop. This is a direct result of having dependencies amongst outputs within a recurrence. Due to this cyclic dependency, placing this statement into the loop  is not possible without a prescient guarantee about what outputs have been calculated for each iteration and location in the loop. These guarantees are made in the form of speculative inductive assumptions. When these assumptions are made, readiness markers encoding these guarantees are placed into the loop nest.

An assumption guarantees that for a given loop iteration variable $i$, the subsequence of output elements that can be considered as already calculated at the start of iteration $i$ is a function of $i$. In the context of an $ijk$ Cholesky decomposition, this could mean that at the beginning of iteration $i$ of the $i$ loop, all rows of the output $L$ up to row $i$ (exclusively) are already calculated. Each assumption can be thought of as a strong inductive hypothesis. In this hypothesis, let $A_{I}$ be a TensorAccess from a fragment's left-hand side, in which $A$ is a tensor and $I$ is an ordered list of indices. Let $v$ be an index variable such that $v\in I$. Let $J$ be another list of index variables, and let $i$ represent the \emph{index of an index variable} in $J$ or $I$. Elements in $J$ or $I$ can use an indexing colon, where a single colon "$:$" represents all potential indices in a dimension, and "$:v$" represents all potential indices less than $v$.

If an inductive assumption is made over $v$, then at the beginning of iteration $v$ in the loop over $v$, an additional set of output TensorAccesses $A_{J}$ can now be considered already computed. Here, $J$ is defined as:
$$
J_i=\begin{cases}
		\; :v & \text{if $I_i=v$}\\
            \; I_i & \text{if an assumption already exists over 
 a parent loop over $I_i$  } \\
            \; : & \text{otherwise} \\
           
		 \end{cases}
$$

The strong inductive hypothesis guarantees that at the beginning of loop iteration $v$, all slices of $A$ until the $v$th slice are already computed. As with strong induction, this hypothesis can only become true if all statements computing values in the $v$th slice of $A$ are placed within iteration $v$ of this loop, such that the $v$th iteration fully calculates the $v$th slice. In RIN, this corresponds to converting the for loop over $v$ to a forall loop. This conversion implies three things:
\begin{enumerate}
    \item The guarantee that all slices of A until slice $v$ will be fully calculated when iteration $v$ of that loop begins. 

    \item The restriction that the current loop over $v$ must be the last loop over $v$ that writes to slice $v$ of $A$. Computations writing to $A$ can now only exist within this loop over $v$, or the aforementioned guarantee will be broken, rendering the inductive hypothesis false.

    \item The introduction of a loop-carried dependency in this loop with respect to $A$, which will affect whether the loop can be parallelized
\end{enumerate}

For example, as shown in ~\ref{fig:compareCholDiagram} in an $ijk$ Cholesky decomposition computing $L_{ij}$, an assumption over $i$ would indicate that all rows of L up to row $i$ are already computed at the start of iteration $i$ in the loop over $i$. Using the formula for $J_i$, we now consider $L(:i\;,\; :)$ as ready in this loop. The algorithm then places a readiness marker $//ready \; L(:i\;,\; :)$ in the first location within the loop over $i$. For the current fragment we are trying to place $L1_{ij} \mathrel{+}= L_{ik}L_{jk}$, the guarantee that all rows up to $i$ are calculated satisfies the dependency over $L_{jk}$, an element in row $j$. The fragment's constraint guarantees that $j<i$, so $L_{jk}$ can now be considered as already computed. The algorithm is then forced to place all statements calculating row $i$ of L within this loop iteration $i$. This is shown by the RIN in the top row of Figure~\ref{fig:cholGen}, which illustrates what happens during this assumption. 

Figure~\ref{fig:cholGen} also illustrates how after an initial assumption is made about $i$ in the $ijk$ Cholesky RIN, an additional, nested assumption is made over $j$, the middle loop in the nest. The formula for $J_i$ indicates that this nested assumption inherits the outer assumption over $i$. The algorithm then places a readiness marker for $//ready \; L(i\;,\;:j)$ at the beginning of iteration $j$ for loop $j$. This implies that at the beginning of iteration $j$ in the loop over $j$, all elements in row $i$ until the $j$th element are already computed . For the current fragment we are trying to place $L1_{ij} \mathrel{+}= L_{ik}L_{jk}$, the constraint guarantees that $k<j$, so the dependency $L_{ik}$ can now be considered as already computed here. As the other dependency $L_{ik}$ was satisfied by the first assumption over $i$, both dependencies are now satisfied in this loop over $j$, allowing the statement to be placed here as shown in Figure ~\ref{fig:cholGen}.

A similar sequence of events will occur for all other loop orderings. In a $jik$ Cholesky ordering where the outer loop iterates over $j$, an assumption over $j$  would guarantee that at the beginning of iteration $j$ in loop $j$, all \emph{columns} of output $L$ up to column $j$ are already computed. This assumption gives a guarantee about columns because $j$ indexes the column dimension in $L_{ij}$.

If the algorithm succeeds, then the inductive assumption is justified. We can apply these speculations to loops over index variables in the order given by the loop ordering, until the number of assumptions made matches the number of dimensions of the highest dimensional output. In this case, once the two assumptions regarding loops over $i$ and $j$ are made, we cannot make anymore, since $L$ is only two dimensional. In the Cholesky example, when the algorithm has placed all fragments in the RIN, iteration $i$ now fully computes row $i$ of $L$ and the corresponding output $L$ is considered fully calculated, as shown in the bottom row of Figure~\ref{fig:cholGen}.

\begin{figure}

\begin{minipage}[b]{0.45\textwidth}
\begin{lstlisting}[numbers=none, basicstyle=\footnotesize]
for_all i<N
  for_all j<i
    for_all k<j 
\end{lstlisting}
\subcaption{Empty skeleton\vspace{1.2em}}
\end{minipage}
\hfill
\begin{minipage}[b]{0.45\textwidth}
\begin{lstlisting}[numbers=none, basicstyle=\footnotesize]
for i<N
  //L(:i,:) ready
  for j<i
    //L(i,:j) ready
    forall k<j
      L1(i,j) += L(i,k)*L(j,k)
    //L1(i,j) ready
\end{lstlisting}
\subcaption{Inductive assumptions made over $i$ and $j$, allowing placement of $L1(i,j)$}
\end{minipage}
\vspace{1em}

\begin{minipage}[b]{0.45\textwidth}
\begin{lstlisting}[numbers=none, basicstyle=\footnotesize]
for i<N
  //L(:i,:) ready
  for j<i
    //L(i,:j) ready
    forall k<j
      L1(i,j) += L(i,k)*L(j,k)
    //L1(i,j) ready
    L(i,j) = (A(i,j)-L1(i,j)) / L(j,j)
    //L(i,j) ready
\end{lstlisting}
\subcaption{Place $L(i,j)$}
\end{minipage}
\hfill
\begin{minipage}[b]{0.45\textwidth}
\begin{lstlisting}[numbers=none, basicstyle=\footnotesize]
for i<N
  //L(:i,:) ready
  for j<i
    //L(i,:j) ready
    forall k<j
      L1(i,j) += L(i,k)*L(j,k)
    //L1(i,j) ready
    L(i,j) = (A(i,j)-L1(i,j)) / L(j,j)
    //L(i,j) ready
    L1(i,i) += L(i,j)*L(i,j)
    //L1(i,i) ready
  L(i,i) = sqrt(A(i,i)-L1(i,i))
\end{lstlisting}
\subcaption{Place remaining fragments}
\end{minipage}

\caption{Steps of the placement algorithm for generating RIN of an $ijk$ Cholesky decomposition.
}
\label{fig:cholGen}
\end{figure}

\subsection{Variable Substitutions}
\label{sec:subs}
Intuitively, a recurrence statement that uses an index variable $i$ should be placed in a loop over $i$ so that the variable is in scope. For recurrences, however, this is actually too a stringent restriction that at best will suppress loop fusion and at worst will prevent correct code generation. In many cases, two index variables with different names (e.g. $i$, $j$) may be logically interchangeable, meaning a recurrence statement that uses $i$ can be rewritten in terms of $j$, allowing it to be placed within a loop over $j$. For example, assume the user defines two recurrences, an identity recurrence $X_i = Y_i : i<N$ defined over $i$ and another identity recurrence  $Y_j = Z_j  : j<N$ defined over $j$. For a given loop ordering of the single variable $i$, a naive greedy placement algorithm would  generate a loop over $i$, note the 2nd fragment is defined solely over $j$, and naively deduce this fragment defined over $j$ cannot be placed in the $i$ loop. However, the 2nd recurrence be safely rewritten as $Y_i = Z_i  : i<N$, as $i$'s iteration space of [$0$, $N$) in the current RIN loop is equivalent to $j$'s iteration space in the original recurrence. In this case the two variables are considered \emph{isomorphic}, and the 2nd recurrence should be rewritten so it can be placed within the the $i$ loop.

A recurrence lowering algorithm should thus have a way to determine if variables can be interchanged when they should be interchanged. Our algorithm can substitute index variables in minimal fragments' TensorAccesses if the new variable $v$ is \emph{isomorphic} to an existing variable $u$. For a particular fragment, two variables $u$ and $v$ are isomorphic at a program location $p$ if the $v$'s lower and upper bound at location $p$ are the same as $u$'s upper and lower bounds in the original minimal fragment recurrence, as determined by the constraints. To determine when variables should be interchanged, let $v$ be the $i$th variable in the loop ordering, $u$ be the jth variable, and $F$ be the fragment (using $u$) we are attempting to place in RIN location $p$, where scope($p$) is the list of all index variables in scope at $p$.

\begin{equation} \label{eq3}
\begin{split}
 S_u = \{ \text{iteration space of $u$ in fragment} \} &\; S_v = \{ \text{iteration space of $v$ in scope($p$)} \}   \\
 u\cong v \iff  inf(S_u) &= inf(S_v) \land sup(S_u)=sup(S_v) \\
 (i<j) \land (u\cong v) \rightarrow \; &\text{replace $u$ with $v$ in $F$}
\end{split}
\end{equation}

Therefore, variable $u$ only needs be replaced with isomorphic variable $v$ if $v$ comes before $u$ in the given loop ordering. Doing this allows fragments to be placed directly into a loop over $v$ if it comes before a loop over $u$ or if a loop over $u$ does not yet exist. These substitutions \emph{normalize} a set of fragments with respect to a schedule and can be done during greedy placement itself. Alternatively, this normalization pass can be executed once after minimal fragment generation and before DAG generation, eliminating the need to rewrite fragments while placing them.

 In the $ijk$ Cholesky decomposition, this can occur when placing the minimal fragment recurrence for $L_{jj} = \sqrt{L1_{jj}}$. Because the fragment calculates a diagonal value where $j=i$, and because the outer loop iterates over $i$ (the first loop in the ordering), it makes sense to redefine $L_{jj}$ in-terms of $i$. This allows the equivalent fragment $L_{ii} = \sqrt{L1_{ii}}$ to be placed in the outer loop over $i$.

This substitution is particularly useful in cases where a summation variable comes before a non-summation variable in the given loop ordering. To illustrate with the simplest possible example, consider the recurrence $S_i = \sqrt{ \sum_{k=0}^{i} S_k} $, where $k$ is a summation variable and $k<i$. The code in Listings~\ref{lst:ik} and~\ref{lst:ki} show RIN for both $ik$ and $ki$ loop orderings. 

\begin{figure}
    \centering
    \begin{minipage}[b]{0.45\textwidth}
        \begin{lstlisting}[language=C++, numbers=none, basicstyle=\footnotesize]
for i<N:
  //S(:i) ready
  forall k<i:
    S1(i) += S(k)
  //S1(i) ready 
  S(i) = sqrt(S1(i))
        \end{lstlisting}
        \vspace{-0.8em}
        \caption{RIN for $ik$ ordering\label{lst:ik}}
    \end{minipage}
    \hfill 
    \begin{minipage}[b]{0.45\textwidth}
        \begin{lstlisting}[language=C++, numbers=none, basicstyle=\footnotesize]
for k<N :
  //S(:k) ready
  //S1(:k+1) ready
  S(k) = sqrt(S1(k))
  //S(k) ready
  forall i>k:
    S1(i) += S(k)
    \end{lstlisting}
    \vspace{-0.8em}
    \caption{RIN for $ki$ ordering\label{lst:ki}}
    \end{minipage}
    \label{fig:kFirst}
\end{figure}

In the left $ik$ listing, an inductive assumption is made over $i$. Once the loop over $k$ completes, $S1(i)$ is considered fully computed, since its minimum fragment recurrence required iterating over all $k<i$, which was accomplished by the inner loop over $k$. The algorithm can then place the statement  $S(i) = sqrt(S1(i))$. 

In the right $ki$ listing, an inductive assumption is made over $k$. Because $i>k$, for outer loop iteration $k$, the inner loop over $i$ writes to all $S1(i)$ where $i>k$. The inductive assumption over $k$ implies that iteration $k$ will be the last iteration in which the inner loop writes to $S1(i=k+1)$, implying $S1(k+1)$ is ready once the inner loop completes. This is equivalent to saying $S1(k)$ is ready at the beginning of the $k$ loop, and we can thus immediately place $S_{k} = \sqrt{S1_{k}}$ there. To handle this, the algorithm makes the substitution $i\rightarrow k$ to rewrite the fragment into $S_{k} = \sqrt{S1_{k}}$.

This above case happens in the $kji$, $kij$, and $ikj$ Cholesky decompositions. This transformation shown in the above listings is actually included as its own transformation in the Symbolic Fractal Analysis \citep{symbolicFractal} compiler framework, which has been used to change loop orderings of the Cholesky decomposition and triangular solve. However, this transformation requires an existing imperative Cholesky code as input. With variable substitutions and RIN, a hardcoded transformation is not necessary, as the above transformation happens automatically. 

\section{Lowering to C}

Our recurrence compiler is written in Python, and the toolchain includes a Python based front-end for defining the recurrence equations, constraints, and schedule. The aforementioned minimal fragment generation, fragment normalization, and greedy lowering occur in this pipeline, which ends with a final pass to lower RIN to C code. This final phase that handles generation of sparse vs dense kernels, in which the loops are customized to iterate over either dense or compressed arrays. 

Our notion of sparsity is adapted from TACO, a tensor algebra compiler that allows users to independently specify a tensor algebra expression (e.g. tensor contractions, element wise tensor operations), the format of the input data (e.g. sparse, dense), and a list of optional scheduling primitives, which are then used to generate native code. Tensor algebra expressions are expressed as operations over a semi-ring, and the semi-ring addition and multiplication are converted into set expressions in which the additions are set unions and the multiplications become set intersections. 

If, for example, the user wants to add two sparse vectors, TACO determines that the program must iterate over the union of A's and B's nonzero locations, as adding a non-zero and zero yields a non-zero. If the user wants to element wise multiply A and B, the program need only iterate over the intersection of A and B's nonzero locations, as multiplying a non-zero and zero yields zero.

As stated, TACO's tensor algebra expressions do not include recurrences in which the left-hand side data structure occurs on the right-hand side. TACO also has no notion of constraints amongst index variables and cannot handle multiple input equations. However, once RIN is generated, due to its similarity to TACO's Concrete Index Notation IR, the compiler can then utilize sparse iteration theory to lower the RIN into C-code iterating over compressed or dense arrays. In this formulation, dense code is merely a subset of sparse code, so the recurrence compiler generates both sparse and dense codes with the same lowering algorithm.  For graph problems, the underlying graph can be encoded as either a sparse or dense matrix, allowing for a graph to be defined via an adjacency matrix (which is dense matrix) or via an adjacency list (which is a compressed-sparse row matrix). 

In RECUMA, for any given recurrence and schedule, the choice of data structure does not affect the correctness of the generated code, even if the the algorithm involves data access patterns for which the data-structure is a "poor fit". RECUMA guarantees that the generated C-code is always correct, and will do-so at the cost of the generated code's performance when necessary.

For example, a compressed-sparse rows (CSR) data structure stores each nonzero entry within a each row, such that the nonzero entries in any particular row are stored consecutively. Such a data structure enables good data locality for algorithms that iterate over rows in the inner-most loop. Conversely, it is possible to use a CSR data-structure with an algorithm that iterates over columns, but only at the expense of the code's performance. Iterating over a particular column in a CSR data-structure requires iterating over all entries (both zero and nonzero) in the column, and randomly accessing each such entry in the column requires an $O(logn)$ binary search within the CSR data structure.

To guarantee correctness, RECUMA will insert expensive binary-searches into the generated code when necessary. If a binary search is necessary, RECUMA will emit a warning that the generated code may exhibit poor performance. Accordingly RECUMA allows a user to quickly determine which data-structure/schedule combinations are viable by ensuring the generated code is free of warnings about $O(logn)$ data accesses.

\section{Optimizations}
RECUMA supports a mix of general program optimizations and domain-specific optimizations targeted towards families of similar recurrences. General techniques like loop fusion and the identification of parallelizable loops are automatic byproducts of the aforementioned code generation process. Other optimizations can also be specified by the user as part of the schedule. The current optimizations described are not an exhaustive list of the many optimizations that could be added. 

\subsection{Timestep Variables}
Several recurrences involve time-stepping, in which $k$ steps are required to convert intermediate results into their final values. In some cases, these intermediate values are useful to store, and in other cases they can be discarded to save memory. Ideally, the user should be able to make this choice, which is possible in RECUMA. To illustrate, we look to the Floyd-Warshall algorithm, an all-pairs shortest paths dynamic programming algorithm. It is inherently dense operates on a 3D tensor, where each dimension is of size $N$ (the number of graph nodes). Its recurrence equation is:

\[ SP_{ijk} = min(  SP_{i,j,k\shortminus1}, SP_{i,k,k\shortminus1}+SP_{k,j, k\shortminus1})\]

$SP_{i,j,k}$ denotes the length of the shortest path between nodes $i$ and $j$, where one can only travel through nodes $0$ through $k$ inclusively. In the generated code, the outer loop iterates over $k$, which acts as a variable denoting the current timestep. The third dimension of $SP$ records the tentative shortest paths recorded at timestep $k$. Each slice $SP(:,:,k)$ denotes the distance between all nodes when only being allowed to travel using nodes $0$ through $k$. The corresponding program input is: 
\begin{lstlisting}[language=Python, numbers=none , frame=none]
rec = "SP(i,j,k) = min(SP(i,j,k), SP(i,k,k-1)+SP(k,j,k-1)) : [0<=j, j<i, i<N]"
storage["SP"] = [Dense(0), Dense(1), Dense(2)]
schedule["ordering"] = "kij"
schedule["removeTimestepDim"] = TimestepVar(tensor=SP, var=k)
prog = Program([rec], schedule, storage]
\end{lstlisting}

This creates the following RIN (with all fragments shown on one line):
        \begin{lstlisting}[language=Python, numbers=none, frame=none]
for k<N:
  forall i<N:
    forall j<i
      SP(i,j,k) = min(SP(i,k,k-1)+SP(k,j,k-1)), SP(i,j,k))
        \end{lstlisting}

The user can safely parallelize the inner two loops, as both will be marked as forall loops by the compiler. Alternatively, some users may not care about the intermediate results held in $SP(:, :, :n-1)$, and only need the final matrix $SP(:, : , n-1)$. RECUMA includes an optimization primitive that lets the user remove the dimension of $SP$ corresponding to the timestepping variable, which here is the third dimension of $SP$. The compiler will then remove the third dimension of $SP$ in the RIN and the storage specification. This resulting program uses asymptotically less memory.  To illustrate, using the \texttt{removeTimestepDim} scheduling primitive on the 3D recurrence

\begin{lstlisting}[language=Python, numbers=none, frame=none]
schedule["removeTimestepDim"] = TimestepVar(tensor="SP", var="k")
\end{lstlisting}
 results in RIN that solves the following 2D recurrence:
\[ SP_{ij} = min(  SP_{ij}, SP_{ik}+SP_{kj})\]

The decision to remove the dimension corresponding to the timestep variable includes a trade-off. While the 2nd recurrence uses much less memory, the resulting recurrence can no longer be parallelized, as parallel in-place updates to $SP_{ij}$ will overwrite values of $SP$ that are still needed. This can be resolved with double-buffering, whose implementation is left as future work. 


\subsection{Parallelization}
The distinction between a for loop and forall loop in Recurrence Index Notation informs auto-parallelization. If a loop is forall and operates over solely over dense data structures, the corresponding C loop can be parallelized trivially with an OpenMP pragma. To illustrate, we use  the Viterbi algorithm \citep{viterbi}, a dynamic program for determining the most likely path taken in a hidden Markov process. It is often treated as a graph problem as well, specifically when the set of states are sparsely connected with edges that represent transition probabilities. The algorithm calculates the most likely sequence of hidden states encountered given a sequence of observations. The associated recurrence utilizes three matrices. Given a transition probability matrix $A$ and an emmission probability matrix $B$, the output matrix $V$ can be calculated with the equation below.

\[ V_{ij} = \max_{k=1}^{N} V_{kj-1} A_{ki}B_{ij} \]

In this case, using a $jki$ or $jik$ loop ordering will result in the outer loop over $j$ being a for loop, and the inner loops being forall. This is shown in the RIN below:
\begin{lstlisting}[language=Python,numbers=none,frame=none]
for j<T:
  forall i<N:
    forall k<N:
      V(i,j) = max(V(k, j-1) * A(k,i) * B(i,j), V(i,j))
\end{lstlisting}
Performance results of parallelization are shown in the evaluation. Enabling parallelization for the loop over $k$ can be done with the following command:
\begin{lstlisting}[language=Python, numbers=none, frame=none]
schedule["omp"] = "k"
\end{lstlisting}


\subsection{Loop Fusion}

As shown, the lowering algorithm greedily places each compute statement into the \emph{first} location of a RIN program where the statement's dependencies are satisfied. This implies that the algorithm will always try to fuse the statement into an existing loop before attempting to place it in a subsequent loop. In other words, loop fusion happens automatically in this framework. Furthermore, the algorithm's ability to handle multiple recurrences implies that statements from separately defined recurrences will, when possible, be placed within a single fused loop. This is highly beneficial for performance, as a RECUMA program will never incur the cost of looping over an array twice when one loop will suffice. This is especially important for memory-bound kernels in which the costs of accessing large arrays dominates a kernel's execution time.  

 To illustrate, we show how two sparse triangular solves will be fused together automatically. Given a lower-triangular matrix $L$ and an input vector $B$, the triangular solve will attempt to solve $Lx=b$ with the following recurrence
\[ X_{i} = (B_{i} - \sum_{j=0}^{i} L_{ij} X_{j}) / L_{ii}\] 

It is common to have to solve the same matrix system $L$ for multiple right-hand side $B$ vectors. In a situation that requires performing solves for two such $B$ vectors, the user can provide the following recurrence equations that solve for $X$ and $Y$:

\[ X_{i} = (B1_{i} - \sum_{j=0}^{i} L_{ij} X_{j}) / L_{ii} \hspace{5mm} Y_{i} = (B2_{i} - \sum_{j=0}^{i} L_{ij} Y_{j}) / L_{ii}\] 

Usually, each triangular solve would require a separate doubly-nested loop over $i$ and $j$. However, if the user provides both equations and an $ij$ loop-ordering, RECUMA will generate a single $ij$ doubly-nested loop that calculates both $X$ and $Y$. This performance benefits of using a fused kernel in this example are demonstrated with experiments discussed in Section~\ref{sec:fusion}.

Two recurrences need not be identical (like in the case above) for fusion to occur; fusion can occur between completely different recurrences. To illustrate, the Cholesky decomposition $A=LL^t$, is commonly followed by two triangular solves to solve for $x$ in the equation $Ax = LL^tx=b$. The first triangular solve can be fused into the outer Cholesky loop. To generate such a fused kernel for an $ijk$ loop nest, the user need only provide the recurrence equations for both the triangular solve and the Cholesky decomposition.  The placement algorithm will then generate a loop nest in which the triangular solve to calculate $X_i$ occurs in the outer-loop over $i$ from the $ijk$ Cholesky loop nest.

\subsection{Masks}

Masks are used in sparse problems to represent a tensor's sparsity pattern. In a breadth-first search, which is expressible as a recurrence, masks can be use to denote the subset of nodes in a graph that are of interest to the user, such that the BFS now only needs to iterate over the relevant data. In various matrix solvers, including the triangular solve and Cholesky, LU, and QR decompositions, they are employed similarly, such that optimized kernels need only iterate over the entries of the matrix that are known to be non-zero. The process of precalculating this nonzero pattern is commonly referred to as symbolic factorization and is used as a preprocessing step to the numerical factorization in which the nonzero values are calculated. 

Generation of the mask is not within the scope of this work, as the underlying process is not necessarily based on solving recurrence. The user hence provides the pre-calculated mask, which is often justified due to many sparse problems consisting of repeatedly solving different matrix problems with the same underlying sparsity pattern. Depending on the sparsity pattern, generating and iterating over masks does not always improve performance. It is thus an optimization parameter for sparse recurrences, and we briefly describe the significance of masks for relevant problems.

In a triangular solve, symbolic analysis can be done with a data structure known as an elimination graph. In this process, given a sparse b-vector, the nonzero elements of b are used as roots to a depth-first search of the sparse matrix $L$, in which a nonzero at location $L_{ij}$ represents an edge between nodes $i$ and $j$. The DFS emits a topologically sorted mask denoting the locations of nonzeroes in $X$, even though $X$'s values have not yet been calculated. For a Cholesky decomposition, a related data structure known as the elimination tree can be used to generate a mask of $L$ in linear time. 

The user can optionally provide a mask, along with an indication of whether the mask is CSC or CSR. When possible, the compiler will then lower any appropriate loops in the RIN to iterate over the mask, thus avoiding needles computations whose output is zero. In cases, where the loops need to iterate over both rows and columns, it may be useful to provide both CSC and CSR masks, which is possible. The primitive for doing so is shown below:
\begin{lstlisting}[language=Python,numbers=none, frame=none]
storage["L"] = Storage([Dense(0), Compressed(1)]) #L stored in CSR
storage["L"].addMask(SparseMask([Dense(0), Compressed(1)])  #CSR Mask
storage["L"].addMask(SparseMask([Dense(1), Compressed(0)])  #CSC Mask
\end{lstlisting}

\section{Evaluation}

We evaluate the performance of recurrence implementations generated by RECUMA against hand-written implementations in existing libraries. We also also perform experiments that demonstrate the importance of fusion, parallelism, different loop orderings, and data formats on performance. 

\subsection{Data Format}
The format of the input data used by kernels can be decided by the user. When the formats of the input and output data are amenable to a program's data access patterns, good memory access locality is achieved which improves performance of the generated code. 

Many recurrences, like the Viterbi equation, are used in both dense and sparse settings. We evaluated RECUMA's ability to generate both dense and sparse variants by changing user specification of the data-structures. The left side of Figure ~\ref{fig:vitNew} shows performance of RECUMA's generated Viterbi kernels when tested with both sparse and dense matrix formats while varying the sparsity of the data. As expected, a sparse format leads to significantly better performance as sparsity increases.

Standard libraries, on the other hand, are tied to one or only a few data formats. CXSparse, for example, provides  handwritten kernels for conducting sparse matrix solves, but requires input data be in a Compressed Sparse Columns (CSC) format. Users who have their input data in alternative format incur a performance penalty from converting their data to the expected form. In a situation where input data is given to the users in a Compressed Sparse Row (CSR) format, we show in Table~\ref{tab:triSolve} the cost of converting the relevant matrix from CSR to CSC, the time to the time to calculate a triangular solve in CXSparse (which requires CSC inputs), as well as the time to calculate a CSR triangular solve in a generated RECUMA kernel, in which no format conversion is necessary. The data conversion routines are optimized kernels from Eigen. For several problem sizes, these conversions take up a nontrivial portion of the program’s runtime when compared to the time it takes to run the triangular solve. RECUMA's generality avoids this problem entirely, as a custom kernel can be generated to fit the input data format.

\subsection{Parallelization}

\begin{figure}
\begin{minipage}{\linewidth}

\centering
\begin{minipage}{0.49\linewidth}
\includegraphics[width=\linewidth]{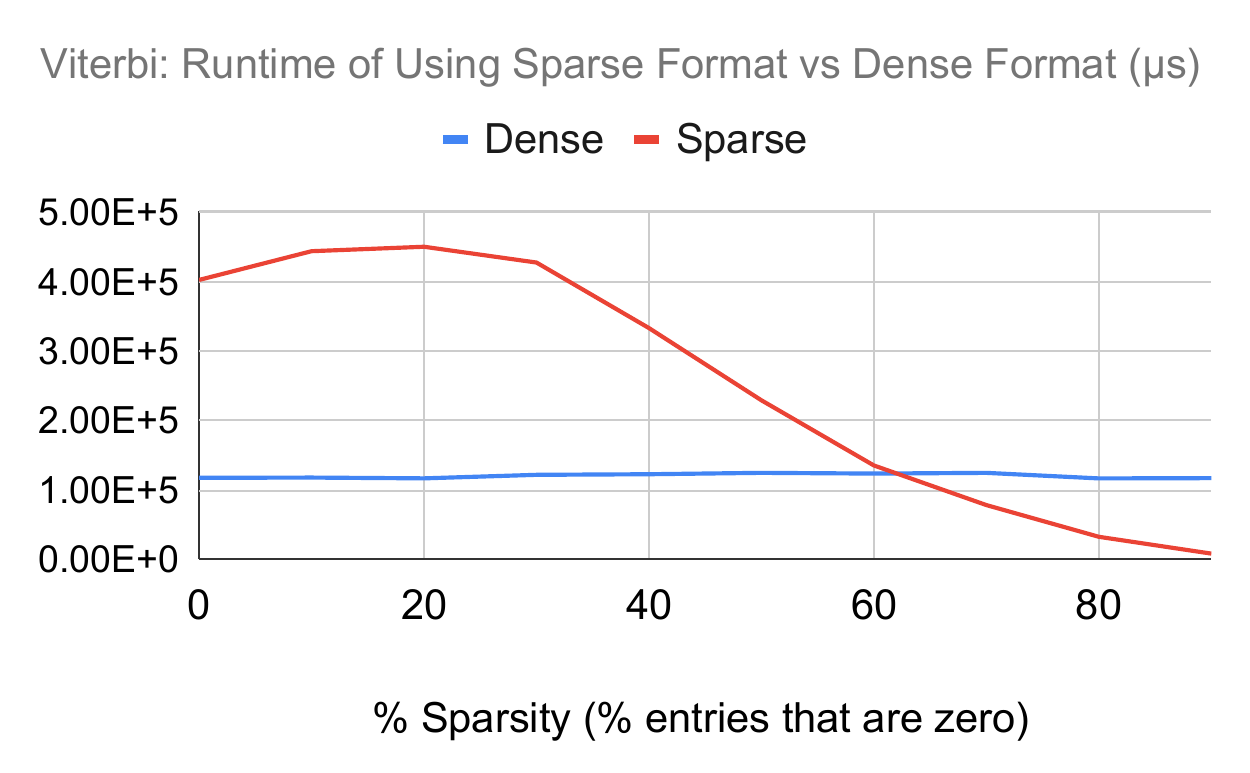}
\subcaption{For a N=1600 Viterbi problem w/ random sparsity pattern, a sparse format is better when sparsity>60\%}
\end{minipage}
\hfill
\begin{minipage}{0.49\linewidth}
\includegraphics[width=\linewidth]{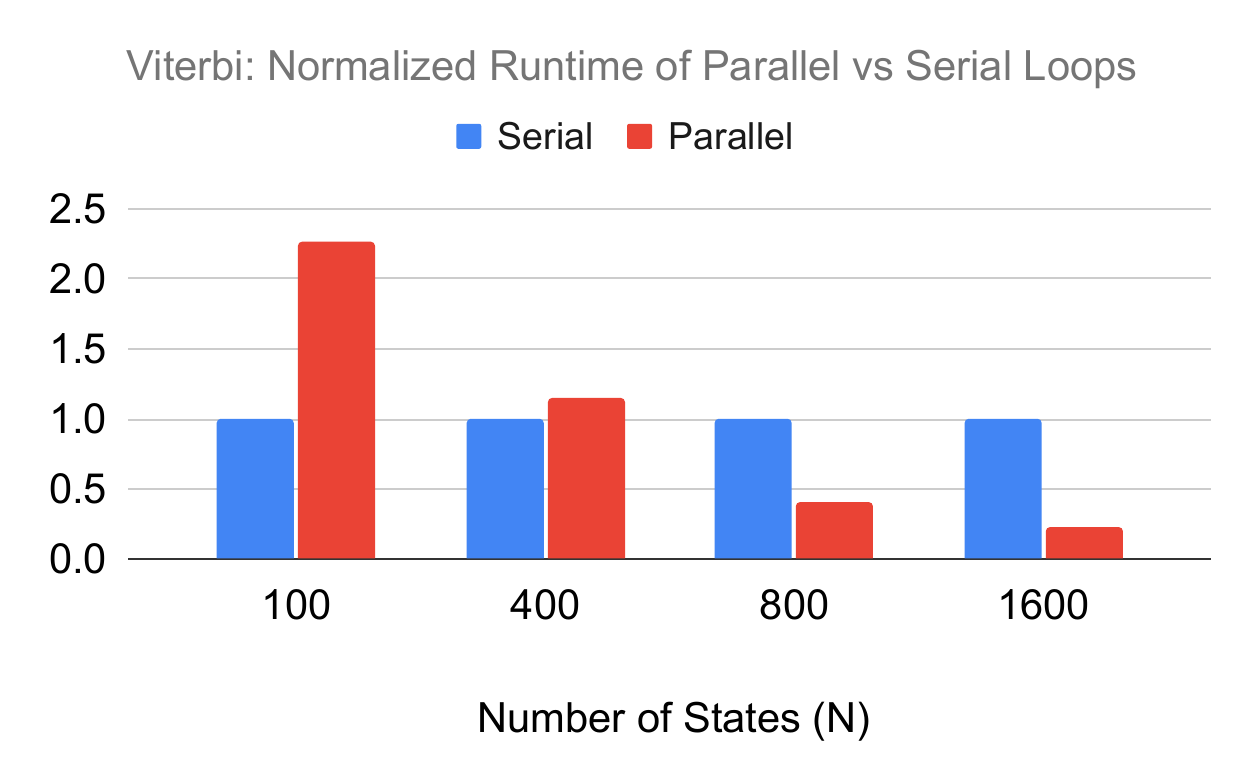}
\subcaption{Viterbi performance improvements of using parallel over serial loops on dense format}
\end{minipage}

\caption{RECUMA Viterbi comparison of (a) dense vs sparse formats and (b) serial vs parallel for dense format} \label{fig:vitNew}
\end{minipage}
\end{figure}

The right side of Figure~\ref{fig:vitNew} displays performance of the RECUMA code when the compiler performs parallelization for the $i$ loop in a $jik$ Viterbi kernel. As expected, parallelization does not help for very small problems where overheads of spawning threads hurt performance but does improve performance as problem size increases. The innermost $k$ loop can also be parallelized, but it is often better to parallelize the $i$ loop to avoid the overheads of spawning many threads for fine-grained workloads. 

\subsection{Loop Fusion}
\label{sec:fusion}
Loop fusion occurs automatically in the RECUMA framework due to the greedy nature of the RIN generation algorithm. Fusing loops avoids the performance penalties of iterating over memory in each loop. To demonstrate the benefits of fusion, we used RECUMA to generate a fused kernel that computes two sparse $ij$ triangular solves for the same matrix $L$ but different $B$ right-hand side vectors. We evaluate the fused performance against that of invoking a separate RECUMA kernel for each triangular solve. The triangular solve is often dominated by the cost of iterating over the matrix $L$. Because the fused kernel only needs to iterate over $L$ once, it performs \emph{nearly} twice as fast as executing two individual triangular solves, as shown in Table~\ref{tab:fusion}.

\begin{table}
  \footnotesize
  \centering
  \caption{Performance comparison of a fused RECUMA kernel to perform an $ij$ sparse triangular solve for two right-handsides vs calling two separate RECUMA sparse triangular solve kernels.}
  \begin{tabular}{|p{3cm}|p{3cm}|p{3cm}|p{3cm}|}
 \hline
 \multicolumn{4}{|c|}{Fused Triangular Solve vs Unfused Triangular Solves (microseconds)} \\
 \hline
 Matrix & Fused runtime & Unfused runtime & Speedup \\
 \hline
1138_bus & 55.85 & 107.4 & 1.92x\\
nasa_10824 & 302.83 & 585.32 & 1.93x \\
bodyy6  & 20539.6 & 40672.4 & 1.98x \\
cbuckle & 3478.72 & 6675.27 & 1.92x \\

 \hline
  \end{tabular}
  \label{tab:fusion}
  \end{table}

\subsection{Loop Ordering}
For a given program, the choice of loop ordering will influence the program's data structure access patterns and therefore have a major effect on performance. Given the flexibility offered by the compiler, a user may select a loop ordering that is amenable to a given input data format. The affect of loop schedules on sparse codes, in which data access patterns are not necessarily regular, is often harder to predict than in dense codes. It is therefore useful to be able to generate different variants and try both. 
Table~\ref{tab:chlsky} shows the performance for $ijk$ and $ikj$ Cholesky schedules, which each perform better on different matrices. On the other hand, for the triangular solve, Table~\ref{tab:triSolve} shows that the $ji$ schedule always performs best on the examples tested. 


\subsection{Comparison with existing libraries}
We evaluate the performance of RECUMA kernels for the sparse Cholesky decomposition, sparse triangular solve, sparse SDDMM, sparse SpMV, dense Floyd-Warshall, dense  Needleman-Wunsch, and dense Gauss-Seidel. SDDMM and SpMv are sparse tensor algebra routines and thus compared TACO kernels. The remaining kernels are compared against existing libraries. Unless noted, all performance metrics were aggregated over 100 trials and conducted on a 40-core Intel Xeon CPU. Graphs showing speedup of RECUMA kernels over existing library kernels are shown in Figure~\ref{fig:all}.

\subsubsection{Triangular Solve}

\begin{table}
\footnotesize
  \centering
  \caption{
  The ratio of the format conversion time against compute time can be significant. None of the kernels shown here use elimination graphs, which are not always useful. Each matrix's Cholesky factor $L$ is used to do the tri. solve, as the original  matrices are SPD but not lower triangular.}
  \begin{tabular}{|p{1.5cm}|p{1.5cm}|p{2cm}|p{2cm}|p{2cm}|p{2cm}|}
 \hline
 \multicolumn{6}{|c|}{Triangular Solve and Format Conversion Runtimes (microseconds)} \\
 \hline
Matrix & nonzeros & CXSparse-ji CSC tri solve & Eigen CSR->CSC conversion & RECUMA-ij CSR tri solve & RECUMA-ji CSR tri solve \\
 \hline
1138_bus &   4054 & 47.371 & 33.824 & 51.25 & 30.767  \\
nasa1824 & 30280 & 201.878 & 114.519 & 181 & 141.614  \\
bodyy6 & 134208 & 21306 & 418.2 & 20277.2 & 18753.9  \\
cbuckle &  676515 & 3291.39 & 1969.58 & 3635.19 & 2946.36 \\
 \hline
  \end{tabular}
  \label{tab:triSolve}
  \end{table}

\begin{figure}
\begin{minipage}{\linewidth}

\centering
\begin{minipage}{0.32\linewidth}
\includegraphics[width=\linewidth]{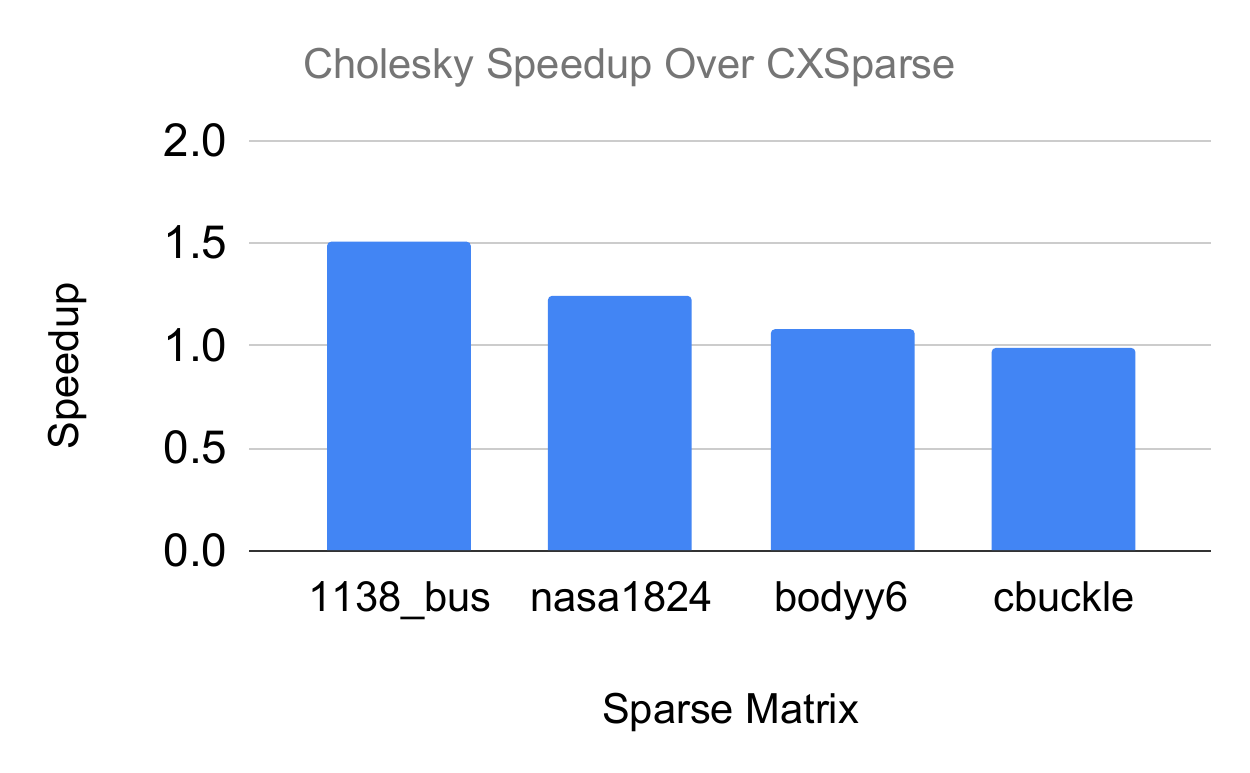}
\end{minipage}
\hfill
\begin{minipage}{0.32\linewidth}
\includegraphics[width=\linewidth]{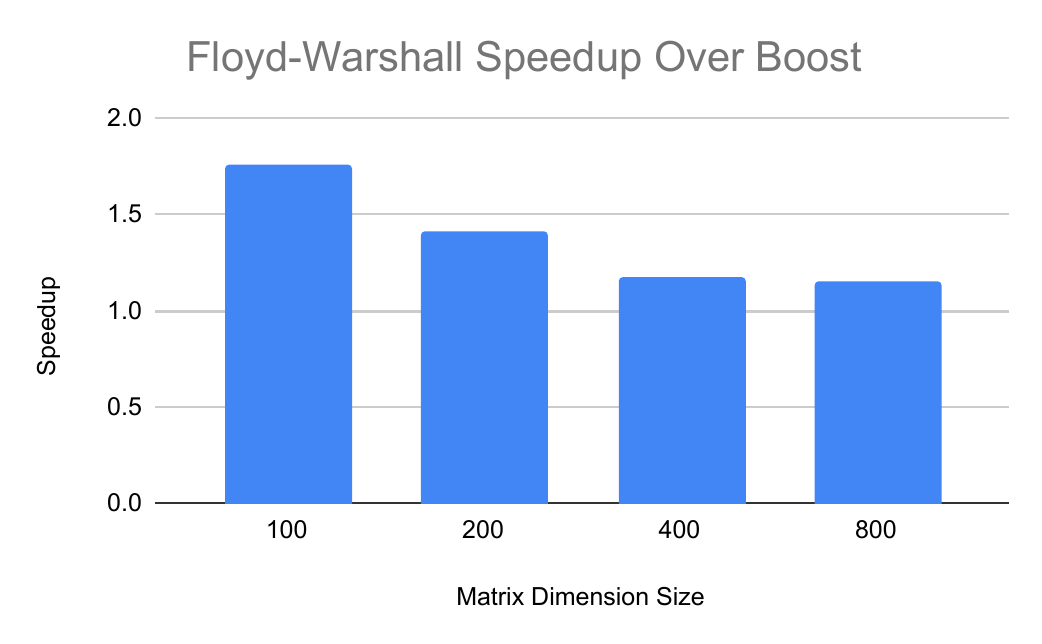}
\end{minipage}
\hfill
\begin{minipage}{0.32\linewidth}
\includegraphics[width=\linewidth]{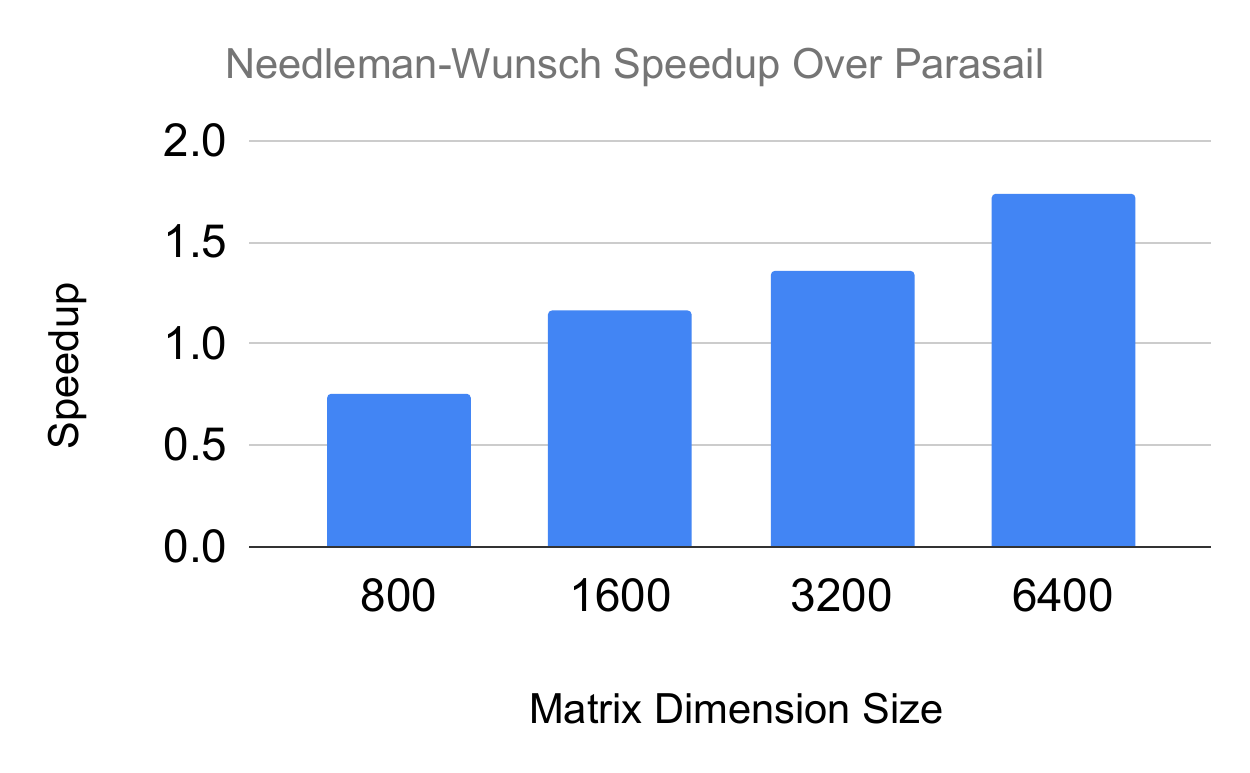}
\end{minipage}

\begin{minipage}{0.32\linewidth}
\includegraphics[width=\linewidth]{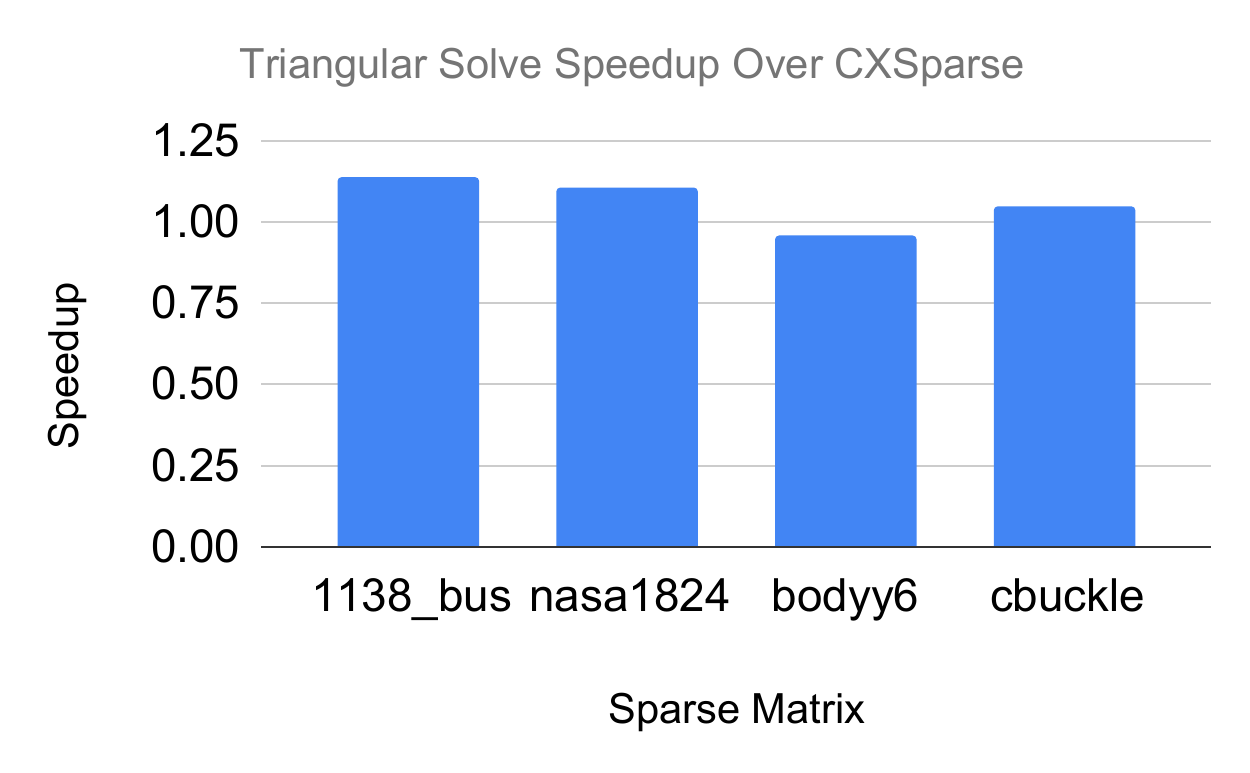}
\end{minipage}
\hfill
\begin{minipage}{0.32\linewidth}
\includegraphics[width=\linewidth]{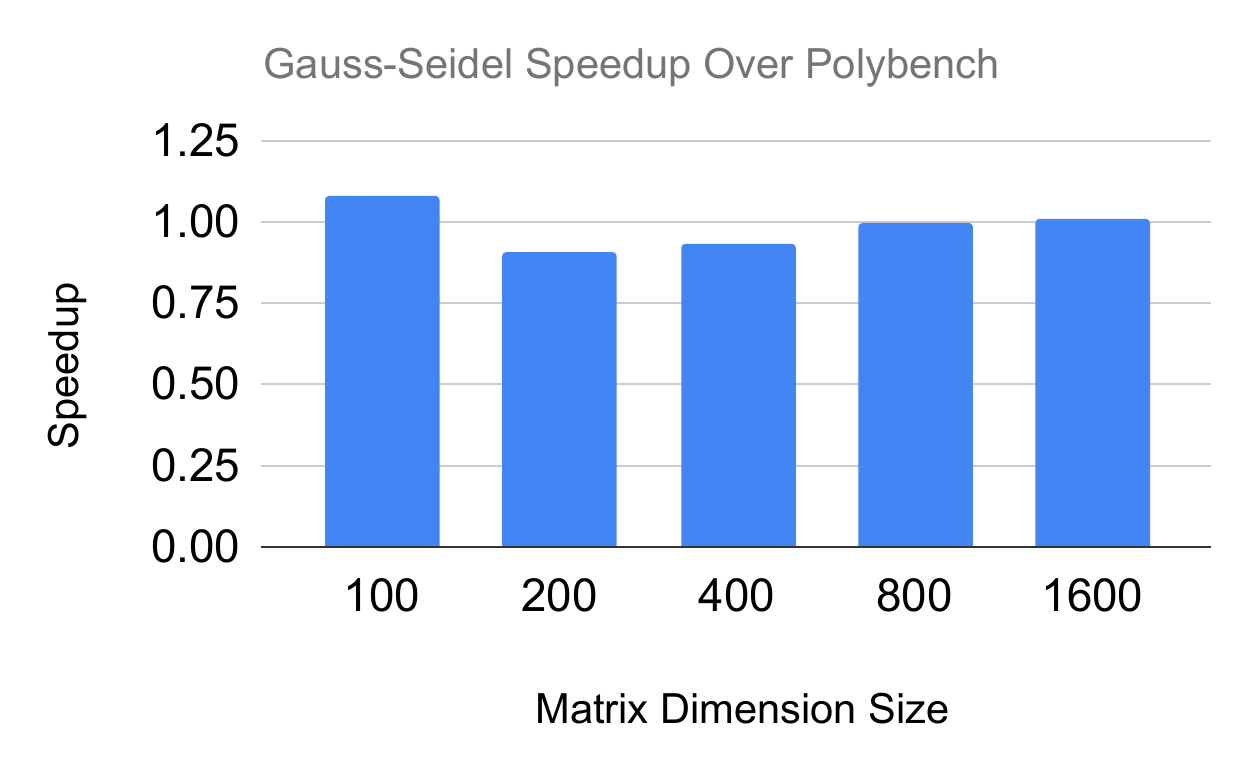}
\end{minipage}
\hfill
\begin{minipage}{0.32\linewidth}
\includegraphics[width=\linewidth]{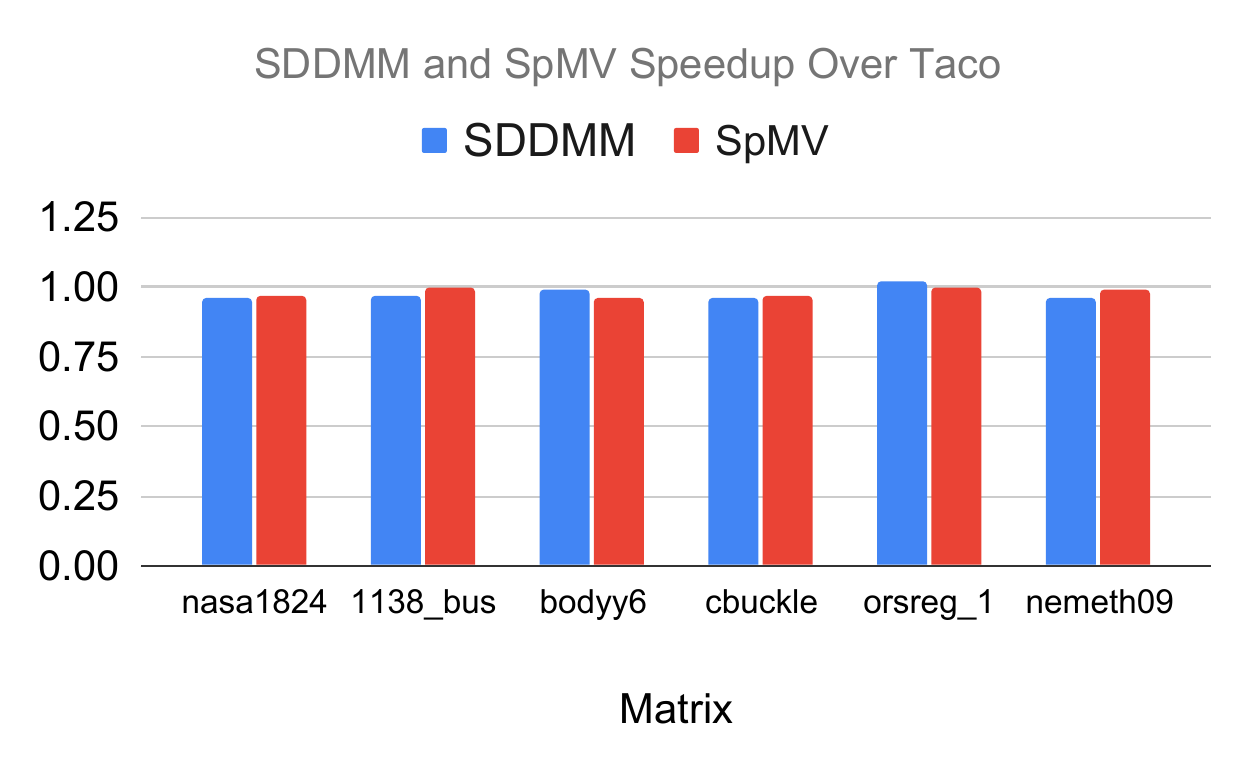}
\end{minipage}


\caption{Speedup of RECUMA-generated kernels over existing libraries. All timings are averaged over 100 trials, except SpMv which uses 500 trials as it executes quickly. Sparse matrices for $ji$ tri solve and $ikj$ Cholesky were chosen via stratified random sampling of SPD matrices from SuiteSparse from the following intervals over \#nonzeros [1-5K],[5K-50K],[50K-500K], [500K-1M]. SDDMM and SpMV matrices include the first four SPD systems used in tri. solve and Cholesky and two additional random non-SPD systems.} \label{fig:all}
\end{minipage}
\end{figure}

We evaluate a RECUMA kernel for performing a $ji$ triangular solve with CSC data structures against CXSparse, whose triangular solve uses the same schedule and data structures. Figure ~\ref{fig:all} shows that their performance is comparable for the matrices tested. 

\begin{table}
   \footnotesize
  \centering
  \caption{Comparison of the $ikj$ Cholesky of RECUMA and the CXSparse Cholesky which also uses an $ikj$ strategy. Matrices were chosen with stratified random sampling from SuiteSparse Matrix collection of SPD matrices to ensure proper spread of nonzero values}
  \begin{tabular}{|p{2cm}|p{2cm}|p{2cm}|p{2cm}|p{2cm}|}
 \hline
 \multicolumn{5}{|c|}{ Cholesky Runtimes (microseconds)} \\
 \hline
 Matrix & nonzeros &  CXSparse ikj perf & RECUMA ikj perf & RECUMA ijk perf \\
 \hline
1138_bus & 4054 & 1976.41 & 1310.05 & 3098.5\\
nasa_10824 & 39208 & 12582.9 & 10123.3 & 21415.1\\
bodyy6 & 134208 & 6415100 & 5916360 & 5765950 \\
cbuckle &  676515 & 441425 & 447210 & 448646\\

 \hline
  \end{tabular}
  \label{tab:chlsky}
  \end{table}

\subsubsection{Cholesky Decomposition}
We generate a RECUMA kernel of an $ikj$ Cholesky decomposition using CSC data structures and compare it against a CXSparse kernel using the same ordering and data structures. Both implementations use a mask, with the main difference being the mask is calculated in the CXSparse Cholesky loop nest, while it is calculated before hand in RECUMA. However, mask calculation performance is almost inconsequential, as it is an $O(n^2)$ algorithm while Cholesky is $O(n^3)$. Speedup plots are shown in Figure~\ref{fig:all}. 


\subsubsection{Floyd-Warshall}
We generated a Floyd-Warshall RECUMA kernel with the timestepping variable optimization to reduce it from a 3D to 2D problem. We benchmark it against a 2D implementation from the Boost graph library. The algorithm is inherently dense, so the connectivity and sparse structure of the underlying matrix is irrelevant to the runtime; only the number graph nodes $N$ matters. The RECUMA vs Boost speedup graph is shown in Figure~\ref{fig:all}, in which RECUMA code outperforms Boost. Boost uses unnecessary C++ features and abstractions that affect performance, while RECUMA's output is simple C code that operates on a C array.

\subsubsection{Needleman-Wunsch}
The Needleman-Wunsch algorithm is commonly used in bioinformatics to align two strings of genomics data. The algorithm operates on a dense 2-dimensional matrix, where each cell is dependent on its left, upper, and upper-left neighbor. To evaluate the code generated by RECUMA, we used the Parasail library as our baseline. As shown in Figure~\ref{fig:all}, for small sizes Parasail outperforms RECUMA, but as we increase the problem size, RECUMA consistently performs Parasail. During compile time, Parasail generates a large suite of Needelman-Wunsch implementations that utilize different vectorization strategies. We tested kernels relevant to the target processor that use the same precision, and show Parasail's best performing kernel in Figure~\ref{fig:all}. We surmise that Parasail's vectorized code is not optimized for the target processor, otherwise Parasail would likely demonstrate superior performance. 

\subsubsection{Gauss-Seidel}
We generated a dense Gauss-Seidel iterative solver, representing an in-place 5-pt stencil. We compare its performance to the Gauss-Seidel solver in the PolyBench benchmarking suite, modified slightly to perform the same computation as our RECUMA kernel. Speedup plots from Figure~\ref{fig:all} show that performance of the RECUMA and PolyBench kernels are nearly identical.

\subsubsection{Tensor Algebra}
We evaluate two tensor algebra kernels: sparse matrix-vector multiply (SpMv) and sampled dense times dense matrix multiply (SDDMM). We evaluated C code generated by RECUMA against C code generated by TACO. The TACO kernels were generated without any additional TACO scheduling primitives. As shown in the speedup plots in Figure~\ref{fig:all} RECUMA expected exhibits nearly equivalent performance to that of TACO for these two kernels.

\section{Related Work}

The Symbolic Fractal Analysis \citep{symbolicFractal} framework has been used to change certain loop orderings of a given dense Cholesky program, but it does not generate an imperative program from a declarative recurrenced. The use of inductive proofs to generate code for different $ijk$ forms of a blocked LU decomposition was explored in the FLAME project \citep{flame}, though it is different in that it operates on recurrences over blocks to generate a program that calculates the decomposition with BLAS calls. \citet{ijk, ijk2} discussed the performance and design of different $ijk$ forms of the Cholesky and LU decompositions. Sympiler \citep{cheshmi2017sympiler} is a compilation framework that can explore the optimization space of sparse Cholesky and triangular solves, generating native code specific to a sparse matrix's sparsity pattern. The CHOLMOD library targets a supernodal \citep{supernodal} form of the sparse Cholesky decomposition, in which subproblems are computed with dense BLAS kernels. Adding supernodal optimizations to a recurrence compiler is left as future work. The supernodal form of the sparse Cholesky decomposition has separately been applied to efficiently solve the all-pairs shortest path graph problem \citep{sao2020supernodal}. For certain loop orderings of matrix decompositions, the multifrontal method \citep{multifrontal} can be used compute the decomposition as a matrix assembly problem. 

Dynamic programming was formalized by \citet{bellman2010dynamic} to address problems in optimization. Gaussian elimination was presented as a type of dynamic program by \citet{lehman1960dynamic}. The process of converting a loop nest that recalculates overlapping subproblems into an asymptotically better dynamic program is addressed in the framework of simplifying reductions \citep{simplifyReduction, simplifyingDepReductions}. \citet{dynamicProgrammingSemiRing} showed that the Viterbi equation, CYK parsing, and certain graph problems are all similar dynamic programming algorithms over different semirings.

Parasail \citep{parasail} can explore the optimization spaces of alignment algorithms and Halide has support for optimizing in-place IIR filters \citep{IIRFilter}. GraphBLAS \citep{graphblas} is a general and performant library that draws upon the dualities between linear algebra and graphs to implement graph algorithms based on recurrence equations. It does not, however, handle dense dynamic programs or solvers. Bellman’s GAP \citep{bellmangap} is another DSL for certain dynamic programs, but like Parasail, is limited to the domain of alignment algorithms. 


The polyhedral model \citep{lamportPoly, FEAUTRIER1991} applies to general programs and was originally applied to finite difference method 
 recurrences \citep{recurrencePolyhedral}. However, the polyhedral model does not handle sparsity and arbitrary data structures. Dyna \citep{dyna} is a DSL that accepts general recurrences like Fibonacci and solves them bottom-up. However, it does not generate a static loop nest and does not generalize across schedules or data structures. It computes outputs by placing every recursive subproblem (i.e. every output element) into a queue, thus requiring expensive runtime analysis to determine which subproblems to solve next. Dyna has similarities to logic programming languages \citep{kowalski2003logic} like PROLOG, which can solve recursive subproblems bottom-up but must handle dependencies at runtime. Dyna can no longer be compiled (as verified by the author), so its exact performance is unknown. 

\section{Conclusion}

We have described a compiler for a language of general recurrences that can express programs across several domains, including dynamic programs, direct matrix solvers, graph algorithms, and tensor algebra. 
The compiler lowers recurrences into imperative loop nests that iterate over dense and/or sparse data structures. Our compiler controls loop ordering and uses induction to manage the dependencies inherent to recurrences. We showed how fusion can be applied to recurrences over sparse data structures, letting us generate code competitive with handwritten libraries. 

Observing that these problems share a theoretical foundation, we envision a shared ecosystem in which optimizations developed for a recurrence algorithm can be easily and readily applied to similar algorithms in other fields, allowing for improvements in program performance and developer productivity to be shared across the respective domains.

\section{Acknowledgements}
This work was supported in part by the Semiconductor Research Corporation (SRC) and DARPA.

\bibliographystyle{ACM-Reference-Format}
\bibliography{main}

\end{document}